\documentclass[preprint]{aastex}
\usepackage{psfig}

\slugcomment{ ApJ Supplements Accepted November 21, 2001}

\shorttitle{UV HST Snapshots of 3CR Radio Sources} 
\shortauthors{Allen et. al.}

\begin{document}


\title{Ultraviolet HST Snapshot Survey of 3CR Radio Source
       Counterparts at Low Redshift}


\author{Mark G. Allen\altaffilmark{1}, 
        William B. Sparks\altaffilmark{1},
        Anton Koekemoer\altaffilmark{1}, 
        Andre R. Martel\altaffilmark{2}, 
        Christopher P. O'Dea\altaffilmark{1}, 
        Stefi A. Baum\altaffilmark{1}, 
        Marco Chiaberge\altaffilmark{1},  
        F. Duccio Macchetto\altaffilmark{1},
        George K. Miley\altaffilmark{3} }
\altaffiltext{1}{Space Telescope Science Institute, Baltimore, MD 21218}
\altaffiltext{2}{Department of Physics and Astronomy, Johns Hopkins University,
3400 N. Charles Street, Baltimore, MD 21218}
\altaffiltext{3}{Leiden Observatory, P.O. Box 9513, NL-2300 RA Leiden, The Netherlands}



\begin{abstract}
We present ultraviolet images of 27 3CR radio galaxies with redshifts
$z<$0.1 that have been imaged with the {\em Space Telescope Imaging
Spectrograph (STIS)} on board the {\em Hubble Space Telescope (HST)}.
The observations employed the NUV-MAMA and broad-band filters with
peak sensitivity at 2200\AA . We find that the UV luminosities show
approximately a factor of 10 to 100 higher dispersion than the
optical.  We compare the UV morphologies with optical $V$- and
$R$-band WFPC2 snapshot survey images. We have found dramatic, complex
and extended ultraviolet emission from radio galaxies {\it even at
zero redshift}.  We find a diverse range of UV morphologies, some
completely divergent from their visual morphology, which are
reminiscent of the chaotic high-$z$ radio galaxies structures seen in
rest-frame UV.  The UV morphologies show regions of star formation,
jets, and possible scattered AGN continuum. The UV emission is generally
not aligned with the radio structure.  We also detect the
diffuse UV emission of the host galaxy.  We propose that these are the
same physical phenomena as observed at high redshift, but on a smaller spatial
scale.
\end{abstract}


\keywords{galaxies: elliptical and lenticular, cD --- galaxies: 
evolution --- galaxies: jets --- galaxies: nuclei --- ultraviolet: 
galaxies --- surveys }

\section{Introduction}

The study of radio galaxies impacts many areas of astrophysics and
cosmology. Typically the hosts of powerful radio sources are massive,
early-type elliptical galaxies that often lie at the heart of
clusters \citep{mat64,yat89,zir96,hill91}.
How these galaxies and clusters are assembled in the early universe
from the initial fluctuations is largely unknown. The detailed physics
of mergers, competing cooling and heating processes and gravitational
aggregation that leads to the massive clusters and giant dominant
central galaxies are currently poorly understood.

Radio galaxies can be seen at all redshifts, thus providing an important,
consistent probe whose systematic characteristics may be traced from
the earliest observable times to the present.  Not only that, but the
fact that they are massive evolving galaxies, often in a most
privileged position at the sites of cosmological structure formation,
makes them especially important in achieving an accurate picture of
the physics and evolution of the universe.

While the relationship between AGN evolution and galaxy evolution
is unknown, there may be links between radio galaxy activity and star 
formation \citep{dey97,bic2000} . 
The epoch of intense star formation in the universe is similar to the
period when the volume density of quasars and radio galaxies 
was orders of magnitude higher than it is today.  Recent X-ray 
observations show directly that
the radio sources themselves interact perhaps in a fundamental way
with the hot coronal intra-cluster gas \citep{smi2001,hard2001} 
that represents one of the most
massive cluster components after the dark matter.

Black hole searches have shown that almost every large elliptical
galaxy harbors a massive black hole \citep{mag98,fer2000}, 
implying all such galaxies go
through an active phase.  The ubiquity of activity in galaxies thus
makes the study of radio galaxies crucial for galaxy evolution and
understanding how this phase affects the star formation, gas and dust
content, and dynamics of the host galaxy.
Investigation of the nuclear environments of radio galaxies is an
important area of AGN physics research, with implications for the triggering,
fueling and evolution of the active nucleus. 

At high redshift ($z>$0.6) the UV rest frame continua of radio galaxies 
are closely aligned with their radio sources \citep{mcc87,cha87}.  The
alignment effect implies that the radio source may play a fundamental
role in the evolution of the galaxy and clusters of galaxies.  The
orientations of lower redshift and lower luminosity radio galaxies
have been studied in detail with claims of both minor and major axis
alignments with respect to the radio lobes \citep{bau89}.

These processes have been well studied at optical wavelengths via the
3CR Imaging Snapshot Surveys conducted with {\em Hubble Space
Telescope} (HST) in Cycles 4-8.  Nearly all the extragalactic radio
sources in the 3CR catalog \citep{ben62a,ben62b,spi85} were observed
in this series of programs.  Impressive results concerning
the prevalence of dust disks and optical jets are described in
\cite{mar98,mar99,leh99,mcc97,dek96}.

Here we describe an observational program to obtain high spatial
resolution HST UV images of nearby powerful 3CR radio galaxies.
We present the results of UV imaging of the low
redshift ($z<0.1$) subset of the sample. These UV images are a major
enhancement to the existing database, and represent the first
systematic high spatial resolution survey of radio galaxies in this
unexplored spectral window.  In particular, using UV imaging we seek
to distinguish the relative roles of gas, dust, jets and star
formation at the present time.
The UV band is well suited to addressing these questions
because of the high sensitivity to the youngest, hottest stars,
and blue synchrotron jets. Also the UV is optimal for detecting 
scattered emission because of 
the blue colour of the illuminating nuclei, and the
high efficiency of scattering at shorter wavelengths.

 A primary
motivation for UV imaging of these nearby 3CR sources is to provide a
zero redshift comparison sample for the extraordinary rest frame UV
morphologies found at high redshift.  In any systematic
study spanning a significant fraction of the age of the
universe it is difficult to disentangle the effects of evolution,
intrinsic power and wavelength of the observation. For the first time
we can empirically characterize the rest frame UV structure of 
radio galaxies at zero redshift.

The details of the observations and data processing are
described in \S~\ref{obs_section}. The UV fluxes and
extinction-corrected luminosities of all the objects are presented in 
\S~\ref{results_section} along with descriptions of the
individual sources and their corresponding
optical and radio properties. In \S~\ref{analysis_section}
the UV and optical luminosities are compared to predictions
from star formation and and ionized gas emission models. 
Discussion is presented in \S~\ref{discussion_section}

\section{Observations and Processing} \label{obs_section}

Table~\ref{observations_table} lists the dates and exposure times 
of the UV snapshot observations.  All the
UV snapshot observations employed the STIS Near Ultraviolet (NUV)
Cs$_{2}$Te Multi-Anode Microchannel Array (MAMA) detector. This
detector has a field of view of 25$\times$25 arcsec$^2$, and a pixel
size of $\sim$0.024 arcseconds.  Most of the objects were observed
with the F25SRF2 filter, and in number of cases the narrower band
F25CN182 filter was used due to the bright object limits of the
NUV-MAMA. The F25SRF2 filter provides a transmission function a
central wavelength of 2320\AA\ and FWHM of 1010\AA\ . The short
wavelength cutoff excludes Geocoronal Lyman-$\alpha$ 1216\AA\
emission. The F25CN182 filter provides medium band width imaging with
pivot wavelength of 1983\AA\ with a FWHM of 630\AA\ .

All the images were processed through the HST pipeline calibration. The
flux calibration of the pipeline is expected to be accurate to 5\%. 
We use the pipeline provided zero-points given by the values of the
PHOTFLAM header keywords which were 5.619$\times 10^{-18}$ for F25SRF2
and 6.144$\times 10^{-17}$ for F25CN182 in units of
ergs~cm$^{-2}$~\AA$^{-1}$. That is the flux in
ergs~s$^{-1}$~cm$^{-2}$~\AA$^{-1}$ is obtained by multiplying PHOTFLAM
by the observed countrate. 

For the 14 objects in our sample with redshift greater 
than z$=0.05$ the F25SRF2 filter 
will include any rest frame Lyman-$\alpha$ 1216\AA\ emission.  
Using IRAF synphot calculations with the assumption of a strong 
emission line spectrum (NGC~1068) we find that  Lyman-$\alpha$ 
can contribute up to 9\% of the total flux in the F25SRF2 bandpass
for these objects.  

The images were rotated to a North up, East left orientation.
The IRAF {\em drizzle} package was used to apply the rotation and
image shifts. The drizzle procedure output sampling was the same as
the input sampling, which leads to slight smoothing of the image
compared to the raw data.  The benefit of using drizzle is that the
flux is preserved, and the loss of resolution is minimized.

The astrometry reported in the image header files is typically
mismatched by up to $\sim1$ arcsecond with respect to the WFPC2 image
of the same object. In order to register the NUV-MAMA image with the
WFPC2 image we applied a shift to the NUV-MAMA image and
used the WFPC2 astrometry. In cases with a bright point source common
to both WFPC2 and NUV-MAMA images a shift was applied to register on
the point source. In other cases the shift was applied to register
common extended image structures. The details of the registrations
applied to each case are described in the notes on the individual
objects. For the present purposes the absolute accuracy of the
astrometry is not important.

Note that in the bright nuclear dominated UV images there is a 
ghost image offset $\sim$1.2\arcsec\ to the left of the object
(in the original unrotated STIS images). This is most likely
due to a refection within the STIS instrument.

\subsection{Background Level and Flux Estimation}

For each object we have measured the total UV and optical emission.
This was done using a set of masks which were defined individually for each
galaxy. Since the images at
different wavelengths were registered spatially and  onto the
same scale, as described above, we use the same suite of masks for the
optical images as well as the UV images.

In order to measure the total UV emission it 
is first necessary to estimate the UV
background level.  The great majority of targets appear to be confined
within the extent of the $25\times 25$~arcsec$^{-2}$ field of view,
although a few show structure extending to the edges of the frame.
Additionally, the NUV MAMA detector has an instrumental effect of
enhanced (dark current) signal around the edges of the detector, see
Figure~\ref{mask_fig}.  
To estimate the background countrate, therefore, we defined a
circular aperture of 800 pixels diameter centered on the MAMA and
excluded the regions of the image exterior to that circle.  Next, we
identified the region of source emission by heavily smoothing the
image with a $\sigma =21$ Gaussian kernel, then thresholding to a
level of 0.1 counts above the background (Figure~\ref{mask_fig2}).  
This corresponds to the
level at which diffuse source emission is starting to become
comparable to the enhanced dark current around the detector edges, see
Figure~\ref{mask_fig}.  The corresponding surface brightness of this minimum
detectable diffuse emission is
$6.7\times 10^{-19}$~ergs~s$^{-1}$~cm$^{-2}$~\AA$^{-1}$arcsec$^{-2}$
for the F25SRF2 filter (the most commonly used) and
$7.3\times 10^{-18}$~ergs~s$^{-1}$~cm$^{-2}$~\AA$^{-1}$arcsec$^{-2}$ for
the lower throughput F25CN182 filter.
The background countrate was then estimated as the average value exterior
to the source defined in this way, and interior to the detector edge mask.
To measure the total UV emission, we summed the counts
within the source region, subtracted the estimated background
and applied the appropriate photometric conversion
as described above.

It is possible that there is a contribution of diffuse extended UV
source emission across the field of view at a surface brightness below
this level.  Figure~\ref{bkd} shows the derived countrates for all the sources.
Sky background on the day side of the orbit contains a significant
contributions from OII air glow emission at 2470\AA\ and OI air glow
at 1302\AA\ . In high-background conditions, the sky background can
dominate the detector background. In average day-side observing
conditions about half the background will be from the sky and half
from detector dark current.  A typical detector plus sky background
level to be expected is $\sim 1.0\times
10^{-3}$~counts~s$^{-1}$~pixel$^{-1}$, with a range $\sim 0.8\times
10^{-3}$~counts~s$^{-1}$~pixel$^{-1}$ to $\sim 1.7\times
10^{-3}$~counts~s$^{-1}$~pixel$^{-1}$.  
As noted above, the STIS pipeline subtracts an estimate of the 
detector background, intended to leave sky and source.
The residual background (ie. dark subtracted) levels we
estimate have a median and standard deviation of $0.37\times 
10^{-3}$~counts~s$^{-1}$~pixel$^{-1}$ and $0.49\times 
10^{-3}$~counts~s$^{-1}$~pixel$^{-1}$
respectively. The residual background shows no correlation with
the pipeline subtracted dark (see Figure~\ref{total_fig}). 
The sky level is expected to be in the
range $6.25\times 10^{-6}$ to $1.2\times 10^{-3}$~counts~s$^{-1}$~pixel$^{-1}$.
As shown in the histogram of background countrates, Figure~\ref{bkd_histogram},
all but three of our residual background measurements are 
within this range. 

In the optical data, there certainly is a contribution of galaxy
emission within the UV background region. We recorded that level also
in order to make color measurements, and return to this in \S~5.1.  In
Figure~\ref{total_fig} we plot the derived UV background versus total
UV flux in order to assess the likely contribution of large scale
diffuse emission.  The lack of any correlation here suggests that such
diffuse emission does not dominate our flux estimates, with the
possible exception of 3C~231 (M~82) where the source fills the entire
field of view.

In Table~\ref{radio_prop_table} the redshift, radio properties, and 
galactic reddening values for each object are tabulated. 

 \section{RESULTS} \label{results_section}

\subsection{Quantitative UV, Optical and Radio properties}

Table~\ref{tfb2_flambda_table} lists the measurements of the total
UV fluxes in units of F$_{\lambda}$(erg~s$^{-1}$~cm$^{-2}$~\AA$^{-1}$)
for all the objects in the sample. In addition we include the corresponding
F702W and F555W fluxes extracted from the same aperture mask. 
Note that these raw flux measurements have not been corrected for 
galactic reddening. 
In Table~\ref{tfb2_lum_table} we list the total UV, F702W and F555W luminosities.
These luminosities were converted from the flux measurements using the
distance derived from the redshift
(H$_{0}=$~75~km~s$^{-1}$~Mpc$^{-1}$) and the filter bandwidth as
specified by the PHOTBW header keywords. The Luminosities also include
a correction for galactic reddening.  The reddening corrections were
calculated using the $E(B-V)$ values from \cite{sch98} (provided via
NED for each individual object's sky position) and the reddening curve
of \cite{car89} as included in SYNPHOT.  To determine reddening
correction factors to the F25SRF2, F25CN182, F702W and F555W fluxes of each
object we used synphot to calculate the inverse transmission of a flat
spectrum with reddening applied with the ebmvx function.

\subsection{Description of Individual Sources} \label{obj_descriptions}

In Figures~\ref{3c29} to \ref{3c465} we compare the UV and optical
images of the sample.
In all the images presented in this section the RA and DEC coordinate axes are 
based on the astrometry reported in the WFPC2 F702W or F555W filter images 
from the optical snapshot survey. The coordinates have been precessed 
to the B1950.0 epoch to facilitate comparisons with published radio maps, 
and to provide consistency with the coordinates shown for the same 
F702W images published in \cite{mar99}.  
In each UV image we also include a bar in the upper right corner 
which indicates the direction of the radio jet axis, and a scale bar
labeled in kpc is included in the lower right corner.

{\em 3C~29}

The UV image of this galaxy is presented in Figure~\ref{3c29}a, and
has been smoothed by a Gaussian with a FWHM of 3 pixels.
The F702W image of this galaxy, shown in Figures~\ref{3c29}b 
shows a very round and smooth light distribution, 
with a relatively faint compact nucleus.
At higher contrast a distorted X-shaped dust lane is visible. This is
seen more clearly in the F555W image 
and in more detail in \cite{spa2000}. The two filaments that cross to 
form the X, intersect at the nucleus of the galaxy and extend 
approximately 1\arcsec\ out from the nucleus. We note that the 
linear feature positioned $\sim$2\arcsec\ NW of the nucleus in 
the F702W image published in \cite{mar99} is an image processing artifact.
The UV image consists of faint extended galaxy emission, visible out to
a 2\arcsec\ radius, and the compact nucleus is detected. The UV image
shows no evidence of the X-shaped dust feature although this is likely
due to the relatively low signal to noise of this observation.
This image represents the first detection of this object in the 
UV as it was undetected with the FOC observation of \citep{zir98}.
Neither the optical nor UV emission show any relation to the radio
jet axis.

{\em 3C~35}

The UV image of this galaxy is presented in Figure~\ref{3c35}a 
and has been smoothed by a Gaussian with a FWHM of 3 pixels
in order to improve the signal to noise of this faint source.
The smoothed image shows a faint compact nucleus surrounded by some 
very low level diffuse emission. There is some indication of 
extended emission extending $\sim1$\arcsec\ south of the nucleus, 
roughly in the direction of the radio jet.  
There is no indication of such an extended structure in the optical image
(Figure~\ref{3c35}b). The asymmetry of optical light distribution is 
due to a 0.5\arcsec\ scale dust lane which curves around the northern
edge of the source. Note that the radio axis is perpendicular to the
dust lane.
  
{\em 3C~40}

The UV image of this galaxy is shown in Figure~\ref{3c40}a, and has been 
smoothed by a Gaussian with FWHM of 3 pixels. The UV image reveals diffuse 
emission over a scale of 3\arcsec. The central region displays a
well defined 2\arcsec\ diameter half-circle of extinction. The straight
edge of the half-circle bisects a weak UV nucleus. The UV morphology 
is well matched to the scale of the elongated dust disk seen in the optical 
image (Figure~\ref{3c40}b). 
(Note that the object is located at the edge of the WF4 chip in the F702W image.)
The UV emission is somewhat irregular in the northern half of the disk,
but there is no apparent relationship to the radio jet axis.

{\em 3C~66B}

The UV image of this galaxy is shown in Figure~\ref{3c66b}a, and has been 
smoothed by a Gaussian with FWHM of 3 pixels. It reveals a 
strong collimated jet of UV emission aligned along the direction of
radio axis.  The prominent jet extends up to 10\arcsec\ from the nucleus
and UV emission is detected continuously along the whole length of the jet.
A strong knot, or kink in the jet occurs at a distance of 3\arcsec\ from 
the nucleus, after which the jet flares and remains more diffuse. 
The jet is faintly visible in the F702W image (Figure~\ref{3c66b}b)
and also in F555W and F814W optical images (not shown here).
This UV jet is an important addition to the 12 or so optical
jets that have been detected in extragalactic radio sources. 
The close relationship of the UV jet to the radio jet morphology, and its
similarity to other known jets clearly indicates that it is synchrotron
emission. A detailed analysis will be presented in \cite{spa2001}.
The compact nucleus seen in the optical images is well detected in the
UV. The small 1\arcsec\ diameter face-on dust disk seen at the
centre of the optical images is not apparent in the UV, although
the host galaxy emission is visible in the UV images.
An additional UV object is detected $\sim6$\arcsec\ north
of the nucleus. This feature has a faint optical counterpart in the 
optical images.

{\em 3C~192}

The UV image of this galaxy is shown in Figure~\ref{3c192}a, and has been 
smoothed by a Gaussian with FWHM of 3 pixels. The UV image displays
a compact core surrounded by diffuse UV emission structures. A 1\arcsec
slightly curved arc, or bar of UV emission is seen on the east side of 
the compact core. A similar slightly larger and more diffuse feature 
is present  0.5\arcsec\ from the core on the opposite side. The lower level 
diffuse UV emission forms a faint spiral structure that is roughly elongated
along the radio jet direction.   
The diffuse UV spiral emission is similar in morphology to 3C~305, 
where the shape is clearly determined by dust
effects, yet the optical F555W
image of 3C~192 shows no counterpart to the diffuse UV emission,
nor does it show any evidence for dust.
(Note that the F702W image from the WFPC2 snapshot survey 
was not pointed correctly and the object fell outside
the WFPC2 field.)

{\em 3C~198}

The UV image of this galaxy is shown in Figure~\ref{3c198}a.
The UV image shows a complex and interesting array of UV 
emission structures. The relatively bright nucleus is
well detected and is surrounded by irregular diffuse emission
that has no counterpart in the optical image (Figure~\ref{3c198}b).
In addition we detect the faint diffuse extended UV emission
of the host galaxy, plus 5 or more compact UV sources 
distributed over 3\arcsec\ scale region. None of these
features appear to be related to the radio jet axis.
The UV image shows a nucleus, surrounded by irregular 
diffuse emission and a number of other point like sources
which have no optical counterparts. A further UV object 
is detected 6\arcsec\ SW of the nucleus, with a faint
counterpart in the optical image. 

{\em 3C~227}

Figure~\ref{3c227}a shows the UV image of 3C~227. This image
is dominated by a strong unresolved UV nucleus. The faint 
feature $\sim$1.2\arcsec\ north of the nucleus is the 
ghost mentioned in \S~\ref{obs_section}.

{\em 3C~231, M~82}

The UV image of 3C~231 is displayed in Figure~\ref{3c231} alongside
the optical F555W image. Note that 3C~231 is not a powerful radio
galaxies, but a famous starburst galaxy. 
The STIS 25\arcsec\ field of view represents
a scale of 330~pc at distance of 3C~231.  The STIS image thus only
covers a small fraction of this large starburst galaxy. For reference,
we include a larger scale F555W image
(Figure~\ref{3c231_big}).
The entire STIS field of view is filled with UV emission showing
spectacular bright and clumpy filamentary structures.
There is also extensive diffuse emission, and compact
clusters of star formation and individual point sources that may
be individual O stars.
There are a number of point source detected in the UV image that
are not detected in the optical image.

{\em 3C 236}

The UV image of 3C~236 is presented in Figure~\ref{3c236}a.  The knot
at the center of the UV image shown here is identified with the
nucleus of the galaxy. Three $\sim0.3$\arcsec\ scale bright knots of
UV emission are visible along an arc which coincides with the SE edge
of the dust lane visible in the optical image (Figure~\ref{3c236}b).
A fainter knot is also visible at the northern extreme of this arc
1.1\arcsec\ NE of the nucleus.  The two brighter UV knots have
counterparts in the optical image.  Weak and diffuse UV emission is
also detected between the knots.  The properties of the UV knots in
this source have been presented in \cite{odea01}. They interpret the
four knots (apart from the nucleus) in an arc along the edge of the
dust lane as star forming regions, and discuss the possible
relationships between the episodic outbursts of the very large radio
source, and the young regions of star formation.

{\em 3C 270 (NGC 4261)}

This galaxy contains a well defined disk of dust and gas which 
surrounds the unresolved nucleus. The structure and dynamics of this
disk has been well studied by  \cite{jaf93,jaf96,fer96}. 
The UV image of 3C~270 (Figure~\ref{3c270}) shows the disk 
clearly defined against the smooth background of the galaxy. 
The nucleus, prominent in the optical, is barely detected
in the UV image.
A faint cone, or jet of UV emission, with apex centred on
the nucleus can be traced 0.5\arcsec\ from the nucleus to 
the eastern edge of the disk. The southern edge of this UV 
emission structure appears to be related to
the dust filament that is apparent in the F547M image \citep{mar2000}. 
The dust filament
is also evident as UV extinction outside the SW edge of the disk. 

The UV emitting cone may be the northern red feature shown
in \cite{mar2000}, in which case scattering from the
dust of a bright nucleus would be implied.
Alternatively, this east-west extension is along the radio jet and
may be synchrotron in nature.

{\em 3C 285}

Figure~\ref{3c285} shows the optical and UV images of 3C~285.
This galaxy has a very chaotic morphology, with multiple large and
small scale irregular dust lanes. There appears to be two main 
dust systems which are elongated perpendicular to one another.
The smaller system, aligned roughly along PA 5 degrees, obscures 
the nucleus. The larger system aligned along PA 330 degrees 
intersects the smaller system at about 2\arcsec\ distance from the nucleus,
and defines a large ridge of extinction which crosses the entire PC
chip (Figure~\ref{3c285}b). 
The UV image reveals clumps of emission which are predominantly 
located along the edges of both the small and large scale dust lanes.
These clumps are suggestive of star formation along the edges of
dust lanes.
Faint diffuse emission is also detected in the UV image, and the
dust lanes are apparent in a smoothed version of the image.

{\em 3C 293}

The UV image of this object (Figure~\ref{3c293}a) 
reveals bright clumps embedded in
filamentary diffuse emission. The large scale structures of the 
filaments are aligned along the same position angle as the dust lanes
seen in the optical image (Figure~\ref{3c293}b). 
Sub-structure within the filaments define strings of small sources
in the shape of arcs and possibly one ring. There are also features
that are reminiscent of bow shock
morphology. 
Most of the UV features do not have optical counterparts. 
The only strong correlation with the optical image is for the 
small (1\arcsec) V-shaped feature which matches the extension of the
optical emission approximately 1\arcsec\ NE of the eastern nucleus.
The UV emission does not show any relationship to the radio jet
and the nuclear peaks are not detected in the UV.

{\em 3C 296}

The UV image of this galaxy, Figure~\ref{3c296}a, shows a well defined, 
smooth dust disk.
The SE side of the dust disk is apparent in the UV image as extinction 
against the background galaxy emission. 

{\em 3C 305}

Figure~\ref{3c305}a reveals extensive UV emission in 3C~305, with a
diffuse fan of emission to the north of the nucleus. The extended
UV emission consists of a complex of knots and filaments where
the radio jet meets the dust disk to the East.
This complex of UV emission also coincides with the major concentration
of [\ion{O}{2}] emission that lies just beyond the end of the radio jet
\citep{jac95}.

{\em 3C 310}

The UV image of 3C~310, Figure~\ref{3c310}a,
displays an elongated central compact
core of emission surrounded by very faint diffuse emission.
The elongation is roughly aligned with the direction of the radio
jet. The optical isophotes however are flattened in the direction of the radio
jet (Figure~\ref{3c310}b).
  
{\em 3C 317}

The UV image of 3C~317 is presented in Figure~\ref{3c317}a. It
displays a relatively bright nucleus, and smooth host galaxy
starlight.  A very blue almost linear filament is evident in the UV
image.  The filament is $\sim$2~kpc in length and is located
$\sim$4~kpc south of the nucleus. \cite{mar2001} show the color of the
filament is most consistent with a recent episode of star formation,
but is not sufficient to represent the inferred amount of cooling gas
if the cooling-flow model is correct.  This source is an example
of the presence of star formation which is {\it not} obviously
correlated with the location of dust.

{\em 3C 321}

The UV image of 3C~321 (Figure~\ref{3c321}a) is spectacular. 
UV emission spans the entire region covered by the two galaxies
seen in the optical image (Figure~\ref{3c321}b) showing bright clumps,
and V-shaped structures. The brightest UV emission
occurs along the northern edge of the dust lane in the southern
galaxy, and the brightest knot is coincident with the partially
obscured nucleus. At least three bright knots can be discerned
in this arc of emission that outlines the edge of the dust lane.
UV emission blobs are detected curving up from the western
end of the dust-lane towards the northern galaxy. From the 
UV image it is impossible to clearly demark the object
into two separate galaxies. 
South of the dust there is a group of UV emission knots embedded in 
diffuse irregular ring feature. Some of the knots in this structure
have corresponding features in the optical image, but others 
do not. (This feature is very suggestive of a 3-D cone with its
axis pointing out from the nucleus of the galaxy.)
In contrast to the relatively smooth and regular optical appearance 
of  the northern galaxy, the UV emission corresponding  to the 
northern galaxy is very irregular. The brightest region, centered on 
the northern
galaxy (northern nucleus) is V-shaped with the apex directed 
toward the nucleus of the southern galaxy. Beyond the flare of
the V-shaped feature, there is more diffuse filamentary UV emission
which has no optical counterpart.
At fainter levels the outline of the northern optical isophotes
of the southern galaxy are apparent in the UV image. 

The radio jet emanates from the nucleus of the southern
galaxy, and follows the PA which remarkably 
passes through the northern neighbor. The spectroscopy
of \cite{rob2000} shows that redshifts of these
galaxies are equal to within 200~km~s$^{-1}$, so these
systems are definitely physically related.

{\em 3C 326}

The UV image of 3C~326 presented in Figure~\ref{3c326}a
reveals a weak compact core, surrounded by faint diffuse emission
against which the dust lane can be seen. The core shows some east-west
elongation, which is most likely due to the dust lane, although this
direction also roughly with the radio jet axis.

Note that there are two candidate galaxies for
the identification of the optical counterpart to the radio source
3C~326. Both sources fall on the WFPC2 images, but only the southern
counterpart was targeted by the STIS snaphot survey. \cite{mar99} also
show the southern galaxy as the radio source counterpart.
 
{\em 3C 338}

The UV emission of 3C~338 is dominated by two bright point sources,
(see Figure~\ref{3c338}a). The brighter southern source coincides
with the nucleus of the galaxy.  We also detect very faint emission from the
underlying galaxy starlight. The prominent optical dust feature 
(Figure~\ref{3c338}b) can just be discerned in a smoothed version of 
the UV image. The northern point source has a faint optical counterpart.

{\em 3C 353}

The UV and optical images of 3C~353 are presented in Figure~\ref{3c353}.
The heavily smoothed UV image shows a weak detection of UV emission.
As noted in \cite{mar99} the optical image shows very round outer
isophotes, with an inner elongated core which roughly aligns with the
direction of the radio jet. The very faint smoothed UV emission
also shows elongation in this direction, although this emission is
only marginally above the noise.
 
{\em 3C 382}

The UV image of 3C~382 (Figure~\ref{3c382}a) 
is dominated by the strong nucleus. The faint object 1.5\arcsec\ 
NW of the nucleus is the ghost described in \S~\ref{obs_section} 

{\em 3C 388}

Figure~\ref{3c388} displays the UV and optical images of 3C~388.  The
UV image shows a compact core plus a weaker UV source located
0.5\arcsec\ SW of the nucleus. This secondary source is aligned with
the direction of the radio jet so we speculate that this feature may
be a UV jet, or is in some way associated with the radio jet. 
The optical image (Figure~\ref{3c388}b) also shows a weak
source next to the compact core, but its position is $\sim0.1$\arcsec\ 
north of the UV jet candidate, and instead corresponds to an even
weaker (highly uncertain) UV source at that location.  None of the
companion galaxies seen in the optical image, and described in
\cite{mar99}, are detected in the UV image.  

{\em 3C 390.3}

A bright UV nucleus dominates the UV image of 3C~390.3 (Figure~\ref{3c390}a).
The faint source south of the nucleus is the ghost image mentioned
in section~\ref{obs_section}. We do not detect the underlying 
host galaxy starlight in this source.

{\em 3C 405 (Cygnus~A)}

The UV and optical images of 3C~405 are presented in Figure~\ref{3c405}.
The UV image displays complex and extended emission with resolved 
bright $\sim0.3$\arcsec\ scale clumps embedded in diffuse emission.
This famous object has been the target of many HST imaging programs
\citep{tad00,tad99,jac98}. The F555W optical image is provided
for reference in Figure~\ref{3c405}b, and F336W, F450W, F662W images 
on the same scale are available in \cite{jac98}. The F555W 
image, while dominated by [\ion{O}{3}] line emission, highlights most of
the major morphological features seen in the optical images.
The UV image presented here, most closely resembles the emission-line
subtracted (ie. continuum) F336W image in \cite{jac98}. Our UV image 
displays the
blue compact features (labeled 1-8 in Figure~5 of their paper) with
much higher contrast and signal to noise. The similarity of the UV image
to these continuum images suggest that the UV emission is also 
dominated by continuum rather than line emission.
\cite{jac98} identify the blue
compact condensations as star formation regions, with the possible 
10\% contribution from nebular emission. They also find that the 
extended diffuse emission is most likely to be scattered radiation 
from an obscured quasar.
The gap through which the radio jet passes is seen in the UV image.
Note that observations of this target are subject to high galactic reddening
of $E(B-V)=0.381$ (see table~\ref{radio_prop_table}) due to its
low galactic latitude.

{\em 3C 449}

The UV image of 3C~449 shown in Figure~\ref{3c449}a reveals a well 
defined $\sim$3\arcsec\ scale disk structure, with significantly 
stronger UV emission in the northern half of the disk. 
The disk is made up of concentric elliptical arcs which have a similar 
size and clumpiness as the optical dust features seen in the dust lane
in the optical image (Figure~\ref{3c449}b).  The position
angle of the UV disk major axis is 168, which is approximately 13 degrees 
offset with respect to the optical dust disk axis.
The western edge of the UV disk matches the curvature of the inside edge
of the dust lane, and there is no UV emission within the dust lane
regions. The UV emission does however resemble the ionized gas 
distribution as shown in \cite{mar2000} with the brightest H$\alpha$
emission in coincidence with the brightest UV emission. 

{\em 3C 452}

A heavily smoothed UV image of 3C~452 is shown in Figure~\ref{3c452}a.
No UV emission is detected in this image. Note that the lower throughput 
F25CN182 filter was used for this observation 
because of the bright star west of the field.

{\em 3C 465 (NGC 7720)}

The UV image of 3C~465 is presented in Figure~\ref{3c465}a.  This
image reveals a compact core of UV emission surrounded by faint
diffuse emission which extends over a scale of approximately 5\arcsec
. The diffuse emission is clearly due to the galaxy starlight.  The
inner region shows a disk structure in absorption against the galaxy
starlight . The morphology is similar to the dust disk seen in 3C~270,
but the contrast between the dust disk and the background galaxy
emission is less in the case of 3C~465. Also, 3C~465 shows a compact
core which is not seen in 3C~270.
The optical F702W image of 3C~465 is shown in Figure~\ref{3c465}b.
It displays a compact central core and an elongated ring of dust
extinction. The dust ring has a projected diameter of $\sim2$\arcsec\ ,
and width of $\sim0.2$\arcsec\ . The outer edge of the optical dust ring
corresponds to the radius of the dust extinction seen in the UV image
\citep{mar2000}.

\subsection{Summary of UV Morphologies and Scales} \label{morph_summary}

In this section we summarize the diverse range of UV
morphologies and emission structures displayed in Figures~\ref{3c29}
to \ref{3c465}. Some objects exhibit similar features 
which allows us to perform a rough categorization into six
groups based on the morphology and scale of the UV emission.
Figure~\ref{sum_fig} shows the categorized objects where
each row of images comprises a single category. 
Figure~\ref{sum_fig} is arranged such that
the first four rows display the groups with extended UV emission 
in order of increasing physical size. 

\subsubsection{Dust Disks}

The objects which display a dust-disk morphology (3C~40, 3C~270,
3C~296, 3C~449 and 3C~465) define a remarkably uniform category
in terms of the scale and physical structure of the of the UV emitting
region. These objects are shown in Figure~\ref{sum_fig} (top row) 
on the same physical scale (2.0 kpc box). 
The three most similar of these objects, 3C~40,
3C~270 and 3C~296, show UV emission that unambiguously comes from the
smooth underlying elliptical host galaxy, with the dust disk apparent in 
absorption against the background UV galaxy light.
The other two objects in this group, 3C~449 and 3C~465 clearly 
have similar scale disk structures, and include contribution from the 
host galaxy starlight, however 
we note that the detailed UV morphologies of these
two dust disks differ from 3C~40, 3C~270, and 3C~296. In the case of 3C~449
the dust disk structure is seen both in emission and absorption.
In 3C~465 the disk is not as regular as 3C~40, 3C~270, or 3C~296
and the source contains a compact central core. We also note that
3C~40 is the only object in this group with no compact core seen
in the optical image.

\subsubsection{Dust Associated Star Formation}

Another striking group of objects are those which exhibit
bright extended UV emission over scales of 5 to 20 kpc.
In a number of cases this large scale emission is clearly
associated with the galaxy dust lane, and in all cases
where this is true, the dust lane displays a very unsettled
and chaotic morphology (3C~285, 3C~293, 3C~236 and 3C~321 ). 
In all these cases the UV emission appears to be 
star formation. This group of objects is 
displayed with a uniform box size of 20~kpc in the third row 
of Figure~\ref{sum_fig}.
We note that the UV morphology in these objects is comparable to 
the rest-frame UV structures seen in optical images of HZRG, and
return to this result in the discussion section.

\subsubsection{Complex Extended UV Emission}

The objects 3C~405, 3C~305, 3C~192 and 3C~198 also display bright,
extended and interesting UV morphology. The emission is
typically extended over a 5~kpc scale in these objects.
While these objects display quite similar features to the
{\it dust associated star formation} group, we assign them to 
a separate group because the UV 
emission is  not so clearly identified as star formation,
and these objects may represent the presence other UV
emission mechanisms. For example, in 3C~405 the bright UV clumps
are most likely due to the young star forming regions 
\citep{jac98}, but the geometry and presence of strong optical
emission lines leads us to expect some contribution from 
for ionized gas and scattered UV emission. 
The UV emission in 3C~305 is closely related to the dust lane 
and may be star formation, however the more diffuse distribution
of the UV emission makes this uncertain, and scattered light from 
the jet may play a role.  
3C~192 is smaller, but
morphologically similar to 3C~305 and may also 
be dominated by scattered light. We loosely categorize these
objects as {\it complex extended} UV emission. The UV images
of the objects in this category are displayed in the fourth row 
of Figure~\ref{sum_fig} with a box size of 10~kpc.

\subsubsection{Compact Core plus Galaxy or Jet Components}

A number of objects show compact cores, in combination with a
significant contribution from the underlying elliptical galaxy
starlight (eg. 3C~310, 3C~35). Some of these objects also
display extra UV emission features such as the prominent UV jet seen
in 3C~66B, and the possible weak jet in 3C~388. Other objects with 
compact cores also exhibit absorption of the UV light by a relatively smooth
dust lane as in 3C~326, or extra UV emitting sources such as the
filament in 3C~317, and the additional UV point source in 3C~338.  We
group all these objects into a category called 
{\it compact core plus galaxy or jet components}. 
Clearly this is a less homogeneous group of objects,
which represent a range of physical mechanisms. The UV images of these
sources are displayed in the second row of Figure~\ref{sum_fig} with a
uniform box size of 5.0~kpc.

\subsubsection{Nuclear Dominated}

Three of the objects in our sample 3C~227, 3C~382 and 3C~390.3, have
very bright UV nuclei at their cores. All of these bright nuclei are
all unresolved at the resolution of HST.  The UV images of this {\it
nuclear dominated} category are displayed in the fifth row of
Figure~\ref{sum_fig} with a box size of 10~kpc.

\subsubsection{UV Host Galaxy and Weak Detections}

The remaining objects consist of the weakly detected UV emission
in 3C~353, and the very faint nucleus plus galaxy host emission
detected in 3C~29. We display this category in the last row of
Figure~\ref{sum_fig} with a box size of 10~kpc.

\section{ANALYSIS} \label{analysis_section}

In this section we make quantitative comparisons of the UV, optical
and radio properties of all the objects in the sample. We investigate
how these properties vary as a function of UV morphology, and compare
the total integrated UV and optical luminosities to theoretical model
predictions of (\romannumeral1)~star formation and evolution, and
(\romannumeral2)~ionized gas line emission.
These comparisons are based on the extinction-corrected, total UV and
optical integrated luminosities from
Table~\ref{tfb2_lum_table}, and the radio properties listed in
Table~\ref{radio_prop_table}.  For this analysis we use
the total integrated optical and
UV luminosities, (rather than separating individual objects into
separate components) since we intend for these results to be useful
for luminosity and colour comparisons with higher redshift radio
galaxies where only the integrated properties are accessible.

First we consider the UV and optical luminosities.  Figure~\ref{m5}a
shows the UV luminosity versus the F702W luminosity, and
Figure~\ref{m5}b shows the UV luminosity versus F555W luminosity.  All
units are given as log$_{10}(L/L_{\sun})$, and selected individual
objects are labeled with their 3C number. Note that 24 of the 27 sample
objects were observed with the F702W filter, 17 were observed with the
F555W filter, and 3 objects were observed only with the F555W filter,
so there is an overlap of 14 objects which are plotted in both
Figures~\ref{m5}a and ~\ref{m5}b.

The most obvious feature in Figure~\ref{m5} is the far higher
dispersion in the UV luminosities as compared to the optical
luminosities. While the optical luminosities range over one order of
magnitude, the UV luminosities range over four orders of magnitude.
Formally, the logarithmic dispersions ($\sigma$) are 0.30 and 0.29 for
F702W and F555W\footnote{3C~231 is excluded in calculation of
F555W$_\sigma$ because it is a Starburst galaxy, and because the
measurements only reflect a small region of the object. See
Sections~\ref{obs_section} and \ref{obj_descriptions}} respectively
and 0.93 for the UV luminosity.  

This result is not surprising given the
extraordinary range in morphology shown in Figures~\ref{3c29} to
\ref{3c465}. Firstly, we expect a far higher dispersion in the 
UV than in the optical because of the wide range of UV emission 
processes indicated by the UV morphologies, compared to the optical
where the emission in most cases is dominated by the old stellar 
population with varying amounts of dust extinction.
Secondly, some of the processes which dominate the UV emission are 
intrinsically more likely to be subject to object-to-object
variations. For example, star formation triggered
by a merger event is a chaotic process, where the amount of
star formation varies greatly with the details of the merger,
(such as gas and dust content of merging systems, and impact
parameter) and moreover the lifetime of the young hot UV-emitting stars
is relatively short. The strength of scattered UV light is intrinsically
variable from object-to-object because of the 
dependence on orientation, and geometry of the scattering medium.
The presence of a bright UV nucleus in some objects 
also contributes significantly to the high UV dispersion
of the total flux. 
Differential dust absorption will also contribute to the 
UV dispersion, but does not appear to be the dominant 
effect because of the clear presence of UV emission mechanisms.

Consider now the UV morphological 
classifications (described in section \ref{morph_summary}) that
are indicated by
different symbols in Figure~\ref{m5}. We find that the UV luminosity is
indeed related to the morphological classifications.
Firstly the three most UV luminous sources (L/L$_{\sun} > 10^{9}$), 
3C~382, 3C~390.3 and 3C~227 are all dominated by very 
bright point sources at their nuclei (L/L$_{\sun} > 10^{9}$). 
The brightness of these sources is obviously due to the AGN out-shining
the host galaxy in the UV by a factor of $\sim100$ or more.
3C~321 also stands out as a very bright source, comparable in
luminosity to the bright UV nuclei (L/L$_{\sun}=1.58\times10^9$), 
yet this source is does not show a strong nucleus, rather it is
dominated by a bright arc of emission along the edge of
this host galaxy's dust lane, with large scale extended 
filamentary and cone shaped diffuse emission. 
3C~321 appears to be the extreme UV bright case of the 
objects classified in our morphological categories
of {\it dust associated star formation} and of 
{\it complex extended}. All the objects in these
two classes are bright in the UV, and there is a clear division 
at the UV luminosity of L/L$_{\sun}=10^8$
between these sources, and the lower luminosity sources
which fall into the other morphological classifications.

The groups of objects classified as 
{\it dust disks} and {\it compact core plus galaxy and jets} are
intermixed in Figure~\ref{m5}, and have UV luminosities 
between  $10^7$ and  $10^8$~L$_{\sun}$. 
The dust disk objects appear to form a very coherent group with
similar UV luminosities, sizes and morphological properties.  As
described in \S~\ref{morph_summary}, all of the dust disk objects
clearly exhibit the same physical structure of a disk surrounding the
central region of the galaxy. The possible differences in the UV
emission processes within this group of objects (host galaxy old
stellar population in 3C~40, 3C~270 and 3C~296, compared to scattering
or ionization in 3C~449 and strong nuclear contribution in 3C~465) is
not apparent in their UV or optical luminosities which show no
systematic differences within the group.

The objects classified as {\it compact core plus galaxy and jets} 
are a less homogeneous group. The UV emission of these
objects includes contributions from the non-thermal 
synchrotron jets, and the nuclei that are
clearly identified in the images. These processes however,
are not made apparent in the total integrated luminosities as 
displayed in Figure~\ref{m5}, where this group of objects 
show a similar range of luminosities as observed for the 
dust disk objects.

The least luminous sources in Figure~\ref{m5} are those 
in which only the host galaxy starlight contributes to the UV
flux. The very faint UV luminosity of 3C~353 is perhaps anomalous,
or due to errors in the background subtraction of this weak source.
The UV and optical luminosities of 3C~231 (M~82) plotted on Figure~\ref{m5} 
represent the total luminosity of the 25\arcsec\ square region 
centred on the nucleus of this starburst galaxy. The UV-optical
colour of this region is similar to the fainter radio galaxy
hosts described here.  

In summary, comparison of the UV and optical luminosities as a function
of their UV morphology described here shows that 
(\romannumeral1) the brightest UV sources are dominated by 
strong UV nuclear emission. (\romannumeral2) Large scale
UV emission features suggestive of star formation and scattering processes
have luminosities well in excess of those galaxies dominated by 
old stellar populations, and in one case 3C~321 the luminosity
is comparable to the nuclear dominated sources. 
(\romannumeral3) 
Objects with compact cores and evidence for UV emission
from jets, have total integrated UV luminosities that are
similar to dust disk objects whose UV emission is dominated by the old
stellar population of the host galaxy.

\subsection{Physical Models: Star Formation and Ionized Gas}

In order to obtain insight into the physical origin of the UV emission
we compare the total integrated UV and optical luminosities to
theoretical model predictions of (\romannumeral1)~star formation and
evolution, and (\romannumeral2)~ionized gas line emission. Here we describe
the models that are to be plotted on the UV versus optical luminosity
diagrams in section ~\ref{model_comparisons}.

Star formation is clearly a major component of UV emission detected in
our images. Using simplified models of the integrated emission we seek
to identify and characterize the old stellar population of the host
elliptical galaxy, and the populations of young UV-bright star forming
regions found in our images.  For this purpose we use a set of stellar
population synthesis models from \cite{bruz93} and, \cite{char01},
which have also recently been described in \cite{odea01} in relation
to star forming regions in 3C~236. In general these models cover a
variety of metallicity, IMF and both single burst and continuous star
formation rates.  Here, we use only the 0.004$Z_{\sun}$ metallicity, Salpeter
IMF models, and calculate the luminosities using the average redshift
of the sample (0.0536).  The model differences over the redshift range of the
sample are insignificant for the model grids as presented here.  Note
that single burst models of ages 7~Gy and older provide colors
expected for an underlying old stellar population.

In some cases the UV emission may come from ionized gas. An overlay of
the STIS and WFPC2 filter bandpasses the spectrum of NGC~1068
Figure~\ref{3c449_bandpass} shows the main UV lines concerned are
\ion{C}{4} $\lambda$1550 and \ion{He}{2}$\lambda$1640 and \ion{C}{3}]
$\lambda$1909. As mentioned above, Lyman-$\alpha$ 1216\AA\ falls
within the UV bandpass for objects with z$>0.05$, however due to the
broad nature of the UV filter, the total flux is still dominated by
the continuum emission (assuming a model shock spectrum, or the
NGC~1068 spectrum).

The ionized gas models used here are shock models
computed with the Mappings III plasma modeling code. Mappings~III is
an updated version of the Mappings~II code described by
\cite{dop95}. Models of shock ionized gas were computed for shock
velocities of $200-1000$ km~s$^{-1}$ with pre-shock density of
1~cm$^{-3}$ and an equipartition magnetic field giving a magnetic
parameter of $B/\sqrt{n} = 3.22$\ $\mu$G\ cm$^{3/2}$.  
For all the Mappings~III models, the UV and optical
fluxes were calculated by passing the model spectra into synphot, and
extracting the fluxes using the appropriate bandpasses. The fluxes
were then converted to units of L/L$_{\sun}$ using the average
redshift of the sample. 

In addition to star forming regions, and ionized gas, other UV
emission processes will also contribute, or may even dominate the
UV flux. In at least one case, 3C~66B, the UV emission is clearly due
to non-thermal synchrotron radiation. Scattered UV emission
may also an important component of the UV emission for the
cases of 3C~321, 3C~305, 3C~449. These processes will be briefly discussed
below (section~\ref{model_comparisons}), and detailed models for 
non-thermal-synchrotron, and scattered emission are deferred to 
future studies of the individual objects.

\subsection{Comparison with models} {\label{model_comparisons}

In this section we compare the UV and optical luminosities to the star
formation and ionized gas emission models described above. For each
set of models we plot the model grid, and the data on diagrams 
of UV luminosity versus F702W luminosity, and of UV luminosity
versus F555W luminosity, where all luminosities are given in units of
log$_{10}(L/L_{\sun})$. 

Figures~\ref{m1} and ~\ref{m2} show overlays of star formation models
calculated for the average redshift of the sample. The first set of 
diagrams displayed in Figure~\ref{m1} show the models for the case
of a {\it constant star formation rate}. As described above, the 
constant star formation models consider the luminosity evolution
of a stellar population in which star formation occurs
continuously, 

The model tracks shown with solid lines in Figure~\ref{m1} trace the
UV and optical luminosities as a function of the age of the stellar 
population. The ages are indicated on the tracks, in units of 
log$_{10}(years)$, up to a maximum age of $2\times10^{10}$ years.
The continuous, constant star formation rate used in 
these models, causes the total luminosity and mass of the modeled 
population  to increase with age, hence the tracks evolve toward the 
upper right corner of the diagrams in Figure~\ref{m1}. The various 
tracks shown, and connected by dashed lines of constant age, 
represent different total masses of the evolved $10^{10}$ year
old populations. The total masses of the models shown are
10$^{13}$ to 10$^{8}$ M$_{\sun}$. These various tracks were simply 
derived from a single {\it constant star formation rate} model, by scaling
the total mass. Since the luminosity scales linearly with
the mass of the model, the dashed lines all have a gradient of 1.0. 
While this scenario may be unphysical due to the fueling requirements 
(input of gas and dust from which stars form)
to maintain a constant star formation rate over these timescales,
we nevertheless include the tracks here to represent one extremum
of star formation.

The majority of the data cannot be described by such models. The four
objects which do fall within the grid are also excluded because they
are strongly dominated by luminous nuclei.  We conclude that constant
star formation models are not applicable.

Figure~\ref{m2} shows model tracks for {\it single burst star
formation}.  These models consider the luminosity evolution of a fixed
mass of stars, in which all the stars are born at the same time.  The
models predict that the UV and optical luminosity of such a population 
stays relatively constant over the first 10$^{6}$ years, then slowly fades. 
Over a typical 10$^{10}$~year lifetime, the UV luminosity decreases by 4 
orders of
magnitude, while the optical luminosity decreases by approximately 1.5
orders of magnitude. Hence the model tracks shown in Figure~\ref{m2} 
evolve in the opposite sense to the previous constant star formation
models, and decrease in luminosity as a function of age. Again the
maximum age considered is  $2\times10^{10}$ years, and the various
tracks shown represent initial population total masses of 
10$^{11}$ to 10$^{7}$ M$_{\sun}$. Note that the total mass of the 
population remains constant with age in the single burst 
scenario, and no attempt is made to include mass loss.
The strong UV evolution shown by these tracks make the UV luminosity 
a sensitive function of the age of the population, and can in principle 
provide an important discriminant for the old versus young populations of stars
that contribute to the UV appearance of the galaxies in our sample.

A significant number of the objects could indeed be described by such
a single burst model.  
All the dust-disk objects are concentrated in the diagram
around the region predicted for an old ($\sim10^{10}$~years) evolved
stellar population of 10$^{10}$ to 10$^{11}$~M$_{\sun}$.  This is
consistent with these sources being dominated by the host elliptical
galaxy star light, with a small correction for dust extinction.

Comparing Figure~\ref{m2}a and \ref{m2}b for consistency between inferred
model parameters using the UV-F702W colours and the UV-F555W colours
we find that the UV bright UV objects common to both the F702W and
F555W diagrams (3C~198, 3C~236, 3C~285 and 3C~305) fall relatively
close to the same location in each diagram with respect to the model
parameters.  The difference is a uniform small shift in the sense that
the models shown in Figure~\ref{m2}b indicate a slightly younger, less
massive population.  This effect however could also be mimicked by
reddening.

Excluding the nuclear dominated sources we find that all the UV bright
sources (categorized as {\it dust associated star formation} and {\it
complex extended}) fall within the single burst
star formation models. This confirms that the colours of those
sources, like 3C~285, 3C~236 and 3C~293 where the morphology suggests
that star formation is the dominant UV process, can be explained by
stellar emission. Clearly these sources are not a single population,
rather mixtures of old and young single burst populations.

To explore this idea in more detail we calculated a mixing
model which includes varying contributions from
young and old populations of stars. In this simplified case
we use luminosities of the old and young extremes of the
single-burst population models described above as the mixing 
components. Figure~\ref{m25} shows the mixing model tracks
for the same total masses as were considered above for the
single burst models. The tick marks along the tracks represent
the luminosity fractions of young to old components. 
The mixing lines show that a small amount of young star formation 
has a significant effect on the UV luminosity. A number of the objects 
are clearly consistent with such a scenario based on the presence of 
star forming morphology, and the match of the integrated colours
expected for such a mixed population. 
  
Figure~\ref{m3} illustrates the UV and optical
luminosities predicted by Mappings~III models of shock ionized gas.
The solid lines of the model grid represent the luminosities 
for shock velocities varying from 100 to 1000~km~s$^{-1}$, with a constant 
physical shock-front area. Both the UV and optical luminosities of shock 
excited gas are strong functions of the shock velocity, increasing by 3 
orders of magnitude over the velocity range considered here. The various
shock model tracks shown in Figure~\ref{m3} have been scaled by the
shock front area in multiples of 10.  The resulting model grid spans only
a narrow region of the parameter space on Figure~\ref{m3}, and due to the 
similar luminosity scaling with shock velocity and shock-front area
the grid is highly redundant. Nevertheless  Figure~\ref{m3} shows that
shock excited gas may generate broadband UV and optical luminosities
in the observed range. The majority of the data points do not fall
within the shock model grid. Those objects that do, 3C~321, 3C~198,
may indeed have contribution from emission lines. 
We also note that there is no systematic difference between the objects in our
sample which contain Lyman Alpha in the bandpass, and those which do
not suggesting that emission lines are not dominant.

Figures~\ref{m12} shows the UV luminosities plotted
against their total radio power at 178 MHz. In Figures~\ref{m12}a the
UV classifications are indicated.  There is no correlation on this
diagram between the large scale, low frequency radio power and UV
luminosity. On the other hand, if we distinguish the objects by their
radio morphological types, FR-I, and FR-II \citep{fan74} then we see
that the two classes are distributed differently on the diagram 
(Figure~\ref{m12}b). 
The FR-I low power sources all have roughly the same UV luminosity,
whereas the FR-II high power sources span the full range in UV
luminosities observed in the sample. The FR-I sources include all the
objects with dust-disk UV morphologies (except 3C~40 which is an FR-II).
3C~29, 3C~338 and 3C~66B are also FR-I sources. The FR-II sources include the 
larger scale dynamic and chaotic UV structures. 
Clearly the more dynamic UV emission processes occur 
preferentially in the FR-II sources, consistent with the
idea that FR-II sources are associated with more disturbed
host galaxy environment consistent with different fuelling of the AGN
in FR-I and FR-II objects \citep{bau95}.

\subsection{Properties of the UV Nuclei}

UV nuclei are clearly an important component of the UV emission
of radio galaxies. Here we draw some broad conclusions about the nature of the
nuclei observed in this sample. We defer the photometry and 
detailed analysis of the optical-UV properties of all the nuclei 
to \cite{chi2001}. 
However, we can compare the integrated luminosities ($\nu L_\nu$) of the
three nuclear dominated galaxies (all of them being BLRG) with those of
radio-loud QSO in the \cite{elvis1994} sample. Limiting ourselves to
the eight sources with z$<$0.3 from their sample, their median luminosity
is $\sim 4 \times 10^{44}$~ergs~s$^{-1}$, 
i.e $\sim 1$ order of magnitude more luminous
than the average of our 3 nuclear dominated galaxies. Although this is
consistent with the current unification scenario, in which BLRG are
believed to be the less luminous counterparts of QSO, without any
further spectral information we cannot establish whether they are
intrinsically fainter or moderately absorbed sources (observed through
the edge of a nuclear obscuring torus, Dennett-Thorpe et al. 2000).

Most of the sources with optical compact cores also show compact cores
in the UV. There are two objects (3C~270 and 3C~296) for which a UV
counterpart is not found for the optical compact core. This may be due
to either a steep nuclear spectrum, or extinction in the nuclear
region.  We note that in both these cases dusty extended structures
are present.

Our observations show that the nuclei of radio galaxies are
indeed bright in the UV and are in most cases visible with greater 
contrast against the background galaxy than in the optical images.
However we do not find any UV compact cores in the sources that 
do not have optical compact cores. This suggests  that if they are 
intrinsically present in these galaxies, as expected in the frame of the 
unified model, they are obscured in both the optical and in the UV by either
nuclear or extended scale structures.

\subsection{Jets}

A strong one-sided UV jet is detected in 3C~66B. 
3C~270 and 3C~388 have possible weak UV jets,
and the nuclei of 3C~192, 3C~310 and 3C~326 
show elongation in the direction of their radio jets.
In order to observe UV emission from synchrotron processes, electron
energies must be high (with corresponding very brief lifetimes), 
and in situ acceleration must be required.
Relativistic beaming effects may play an important role, both
in boosting the intensity of radiation and also in blue-shifting
the synchrotron break frequency which elevates the UV emission.
There may be other examples of synchrotron radiation amongst
the features we observe, however additional observations
and detailed comparisons with radio structures
would be needed, and are beyond the scope of this
paper.

\subsection{Scattering}

We identify 3C~270, 3C~321, 
3C~305, 3C~405, and 3C~449 as candidates for scattered UV
emission  based on their morphology. 
Both 3C~321 and 3C~405 show conical structures which are likely
the result of illumination by a strong hidden nuclear source. 

In 3C~305 and 3C~192 the radio jet is roughly aligned with the
extension of the UV emission, suggesting that in these cases
we may be observing scattered jet photons.

Two of the dust disk like objects, 3C~270 and 3C~449 show
signs of scattered emission. In 3C~270, the UV enhancement
at the western edge of the disk may represent scattered
nuclear continuum, or possibly scattered jet photons.
These two sources are however both FR~1 type objects  
which according to some authors, are not expected to have a a strong (hidden)
source of UV continuum available to be scattered.
\cite{bau95} find that emission-line gas in FR~2 objects are photoionized by
strong nuclear UV continuum, and that FR~1 objects are not.
This apparent lack
of a strong ionizing UV continuum source in FR~1 objects, suggests that 
we may not expect to find significant levels of scattered UV nuclear 
light in the FR~1 objects in our sample. 

\section{Discussion} \label{discussion_section}

As described in the above classifications of our UV images, we find
that the processes of star formation, nuclear emission, scattering,
and extinction by dust, are the basic components which determine the
UV appearance of low redshift radio galaxies. The relationships between
these various components provide clues to the physics, merger history
and evolution of radio galaxies. These same processes are also 
the likely ingredients which determine the appearance of HZRG.

We defer detailed analysis of the multiwavelength properties and
of the individual objects to future papers. Here we briefly discuss
some implications of the star forming regions found in the 3C radio
galaxies, and make a broad comparison of the low redshift UV morphologies
with high redshift radio galaxies.  

We find a progression of UV morphologies from highly chaotic dusty
systems with strong star formation, to smooth regular systems with
nuclear dust-disks that are aligned perpendicularly to the radio jet.
This range of morphologies appeals to a scenario where the central
regions of radio galaxy hosts may evolve from highly chaotic systems,
to more stable regular systems. This idea is linked with the merger
scenario for the triggering and fueling of the active nucleus.  In
this evolutionary scenario the categories of objects identified in
this paper may roughly represent the various stages of the effect of
infalling gas and dust. For example the large scale chaotic dust lanes
of the {\it dust associated star formation} group may represent the
earliest stages of a merger event. The radio axes of these objects
bear no relation to the dust lanes or central structures in these
galaxies, perhaps indicating that these systems have not had enough
time to settle into a stable configuration.  At the other extreme the
{\it dust disk} objects do not show strong evidence for star formation
associated with the dust, and these regular dust disks may represent
the final stable configuration.  In all the dust disks the radio jet
is aligned roughly perpendicularly to the major axis of the disk.

One of the difficulties with such a simplified scenario is that the
timescales for settling of gas is of order 1~Gy after the merger
\citep{gunn79,chr93}, which is longer than the $\sim10^7$ year
duration of a single epoch of radio activity \citep{ale87}. If a
merger resulted in only a single epoch of radio activity, then there
would not be enough time for evolution from a 3C~293 like object into
a settled dust disk within the lifetime of the radio jets. The
evolutionary scenario may however be credible if the onset of radio
activity occured (or re-occurred) at a time after the merger, perhaps dependent
on the precise parameters of the encounter. These parameters may include
merger impact geometry and gas infall timescale, also the gas and dust 
content of the merging systems, and the angular momentum loss rate.  
If the onset of radio activity occurs at random time during a merger
we may be able to account for the coexistence of radio jets and the wide range
of morphologies as described here. Alternatively FR-I sources may
be intriniscally different due to a different low accretion fuelling
rate for the AGN as described in \cite{bau95}.

Estimating the ages of the newly triggered star formation may 
provide a means of dating major merger events in the history of the
host galaxy. Similarly, if star formation is related to the nuclear
activity of the galaxy, the star formation ages will significantly
aid our understanding of these systems as has recently been 
demonstrated by \cite{odea01} for the case of 3C~236 (using the UV
and optical data described in this paper). They showed 
that the young star forming regions may be related to the inner 
2~kpc scale radio source, and that the star formation may reflect the 
radio source history.  Clearly many of the {\it dust associated star
formation} objects show similarities to 3C~236 and we expect that 
more accurate dating of the individual star forming knots will 
shed light on the merger and activity history of these galaxies.

A primary motivation for UV imaging of low redshift radio galaxies is
to provide a zero redshift comparison sample for the extraordinary
rest frame UV morphologies found at high redshift.  High resolution
HST images of high redshift radio galaxies have been described by
\cite{bes97} ( 28 3CR radio galaxies $0.6<z<1.8$) and \cite{pen99} 
(9 distant radio galaxies $2.3<z<3.6$).  In both of these samples the
optical observations sampled the rest frame UV of the radio galaxies,
providing an ideal comparison for the rest frame UV emission of low
redshift galaxies presented here.  In general these objects display a
spectacular range of structures, some but not all are aligned with the
radio structures. The morphology of the emission varies greatly from a
single bright patch to strings of knots stretching along the radio
source. The UV emission of most objects is clumpy and irregular,
consisting of a bright nucleus and fainter components. Some of the
structures may be due to distributed dust.

A common feature in high redshift radio galaxies is that the 
optical and radio structure align, implying a fundamental relationship 
between the
jets, and the host galaxy. \cite{bes96,bes97} and \cite{pen99} find that the UV continuum is generally elongated along the radio source axis, although the
characteristics of the alignment differ from case to case suggesting
that the alignment is not the result of a single physical mechanism.

In general the low redshift radio galaxies do not show the high degree
of alignment as observed at high redshift.
Within the 27 objects presented in this paper we find only 2 objects
show rough alignment of their UV emission with the direction of the
radio jet with some similar characteristics to that seen at high
redshift.  The extended UV emission observed in 3C~321 is the most
obvious case. The conical features are similar to that seen in the
high redshift radio galaxy TX~0828+193 \citep{pen99}. We note however
that 3C~321 is highly complex with strong star formation along the
dust lane which is in general not observed in the high redshift
sources.  3C~305 also displays elongation of the UV emission along the
direction of the radio jet.  3C~305 is similar to MRC~0943-242\citep{pen99} 
in terms of UV emission morphology, and radio source size.  
3C~405 is unique in the sample being by far the most powerful radio 
galaxy, with total luminosity similar to $z\sim1$ radio galaxies.
The UV emission in 3C~405 is clearly related to the axis of the radio
jet in that the UV emission is distributed in the bi-conical structure
obvious in emission-line images of this galaxy. The UV emission consisting
of star forming regions and possible scattered light is however not
elongated in the direction of the radio jet.

The UV images of low redshift ($z<0.1$) radio galaxies presented here 
provide a high quality database for studying many important issues
for our understanding of radio galaxies, and galaxy evolution in general.
Using integrated properties we have shown that it is possible to
characterize the star forming regions around the edges of the chaotic
dust lanes seen in some of these objects. This has implications
for dating the important events in the histories of these systems. 
Detailed studies of the individual objects are expected to provide
insight into the relative roles of gas, dust, jets and star formation
in radio galaxy hosts. In addition these data provide important
constraints for identifying and characterizing the processes of scattering
and ionization, and the spectral energy distributions of the radio
galaxy nuclei and jets.  

\acknowledgments


\clearpage



\figcaption[fig1.ps]{ Full field of the STIS NUV MAMA observation
of 3C~236, smoothed with a  $\sigma =21$ Gaussian kernel. Note the enhanced 
background around the edges of the frame, which is due to residuals from
the dark current subtraction. The inner contour represents the 
source region, defined as 0.1 counts above the background. The background
was measured in the region between the source and the 800 pixel diameter
circle shown. The outer circle was adapted in cases where a possible 
source fell near its edge as in 3C~236 as shown.  \label{mask_fig}}

\figcaption[fig2a.ps,fig2b.ps]{ Registered NUV MAMA and WFPC F702W frames.
This example of 3C~236 shows the UV and optical images
which have been {\it drizzled} onto the same pixel grid
(STIS pixel size) including a N-E rotation and object registration. 
The object masks shown have also been appropriately rotated and
registered in order to extract total fluxes from the same regions
in the UV and optical images.     \label{mask_fig2}}

\psfig{figure=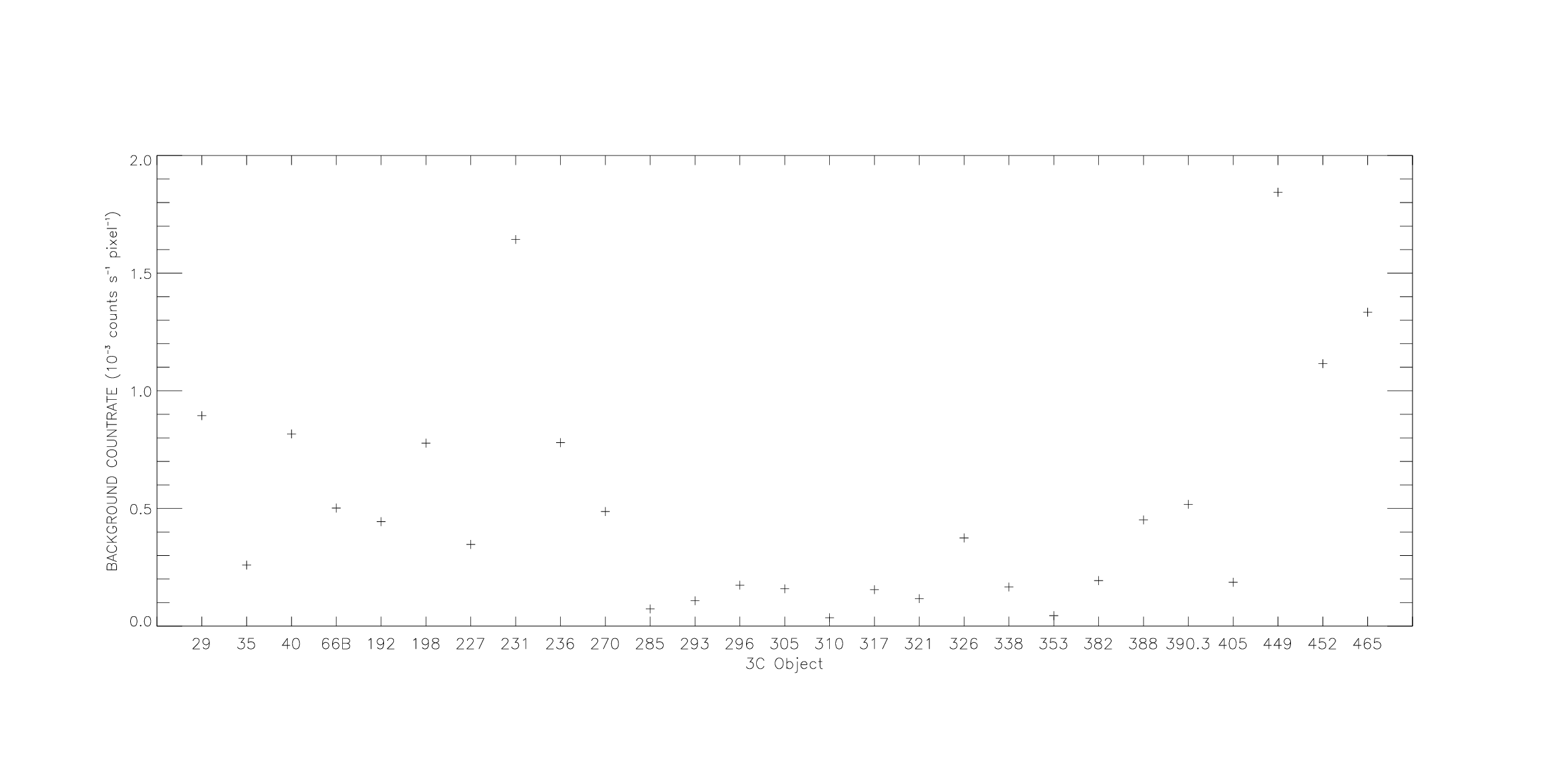,width=5in}
\figcaption[fig3.ps]{ UV Background rate measurements. The UV background
rates measured from the region outside the object region and inside the
800 pixel diameter circle, are shown for each object. \label{bkd}}

\centerline{\hbox{
\psfig{figure=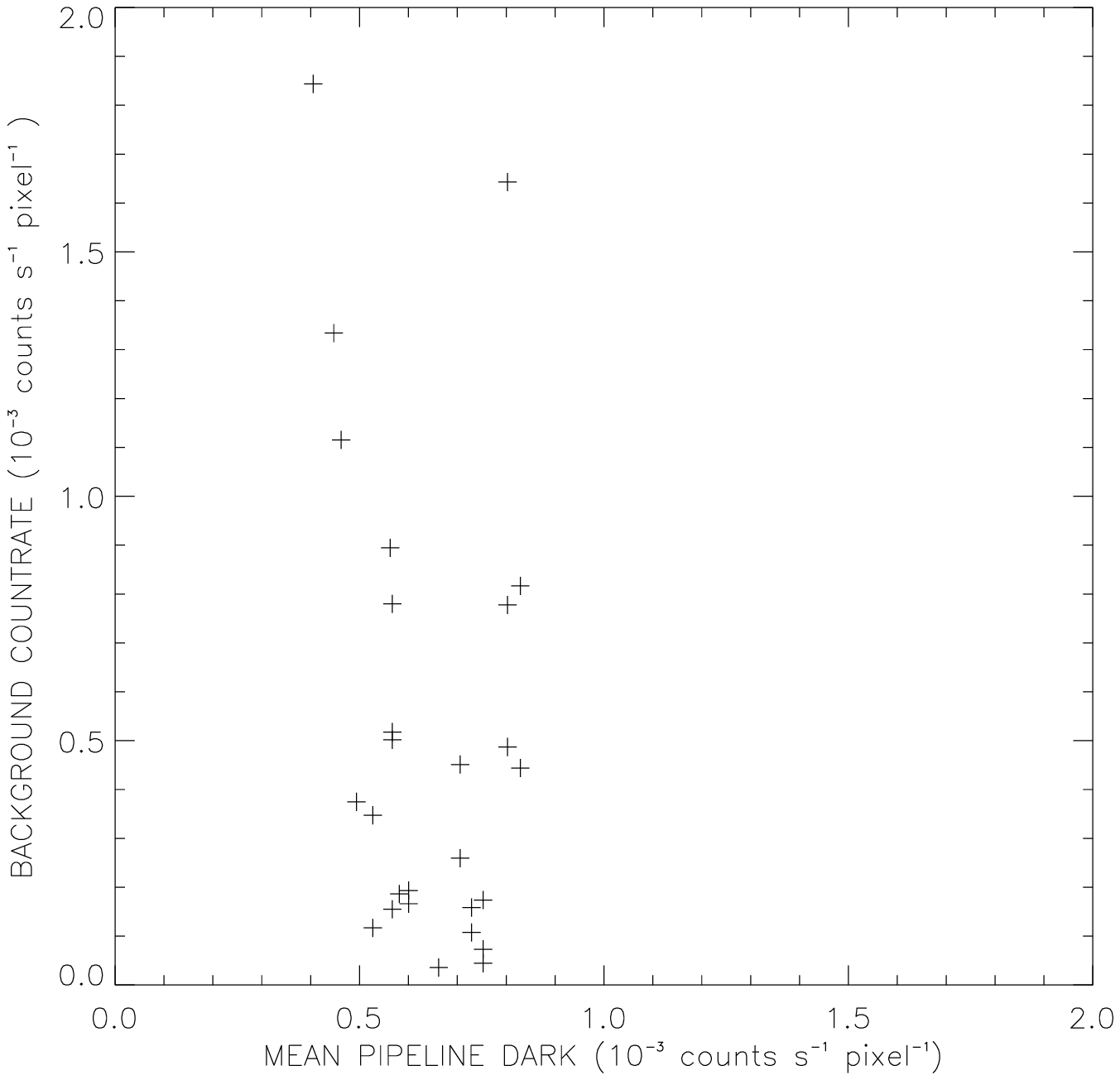,width=3in}
\psfig{figure=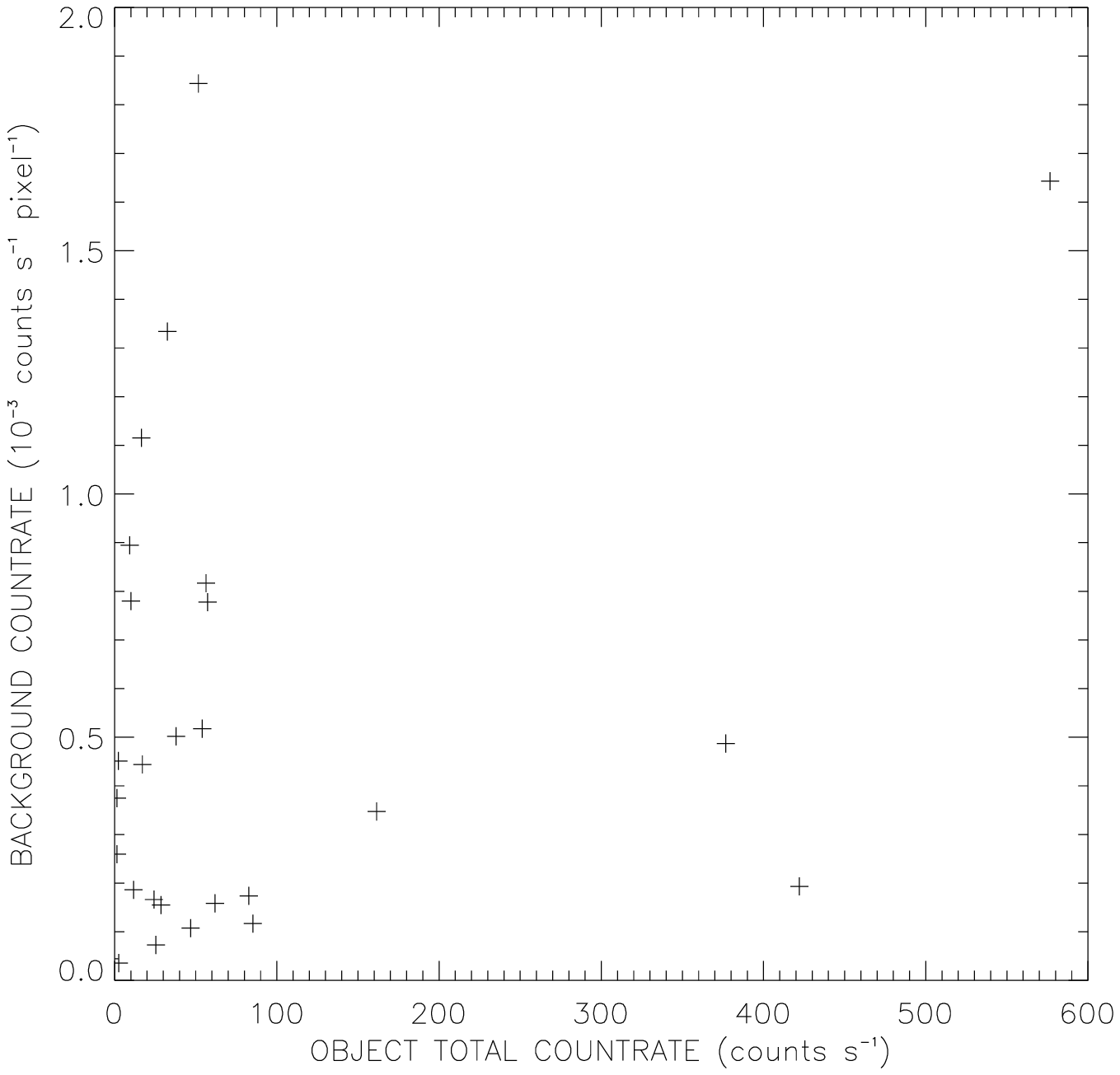,width=3in}}}
\figcaption[fig4a.ps,fig4b.ps]{ Background, Mean Dark and Total Object
Countrates. (a.) The background countrate measured from our
images is shown versus the mean dark countrate that was subtracted in
the pipeline reduction of the images. Note that there is no
correlation between these two countrates. (b.) The background 
countrate measured from our images is shown versus the total object
countrate. There is no correlation of the object countrate rate with the
background with the exception of 3C~231 (M~82) where the background
is clearly overestimated due to emission covering the entire field
of view. The lack of a correlation for the majority of the objects
provides confidence in our background and total flux measurements 
\label{total_fig}}

\psfig{figure=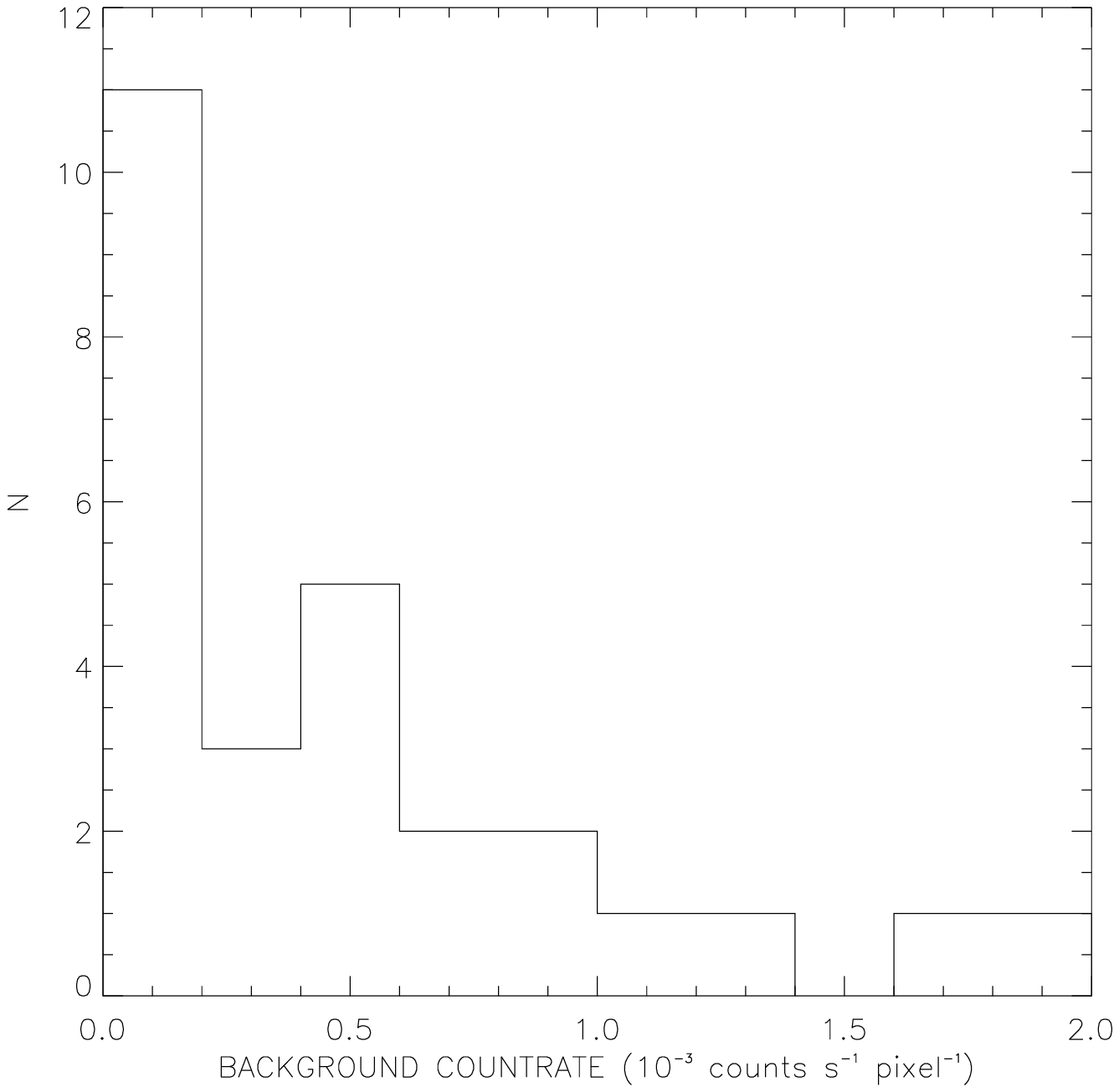,width=5in}
\figcaption[fig5.ps]{Histogram of the UV Background Rates.
The background measurements as described in Section~\ref{obs_section}
are expected to be dominated by the sky background. With the
observational parameters used the expected sky countrate is
$6.25\times 10^{-6}$ to $1.2\times 10^{-3}$~counts~s$^{-1}$~pixel$^{-1}$.
All but three of the objects fall within this range.
 \label{bkd_histogram}}

\psfig{figure=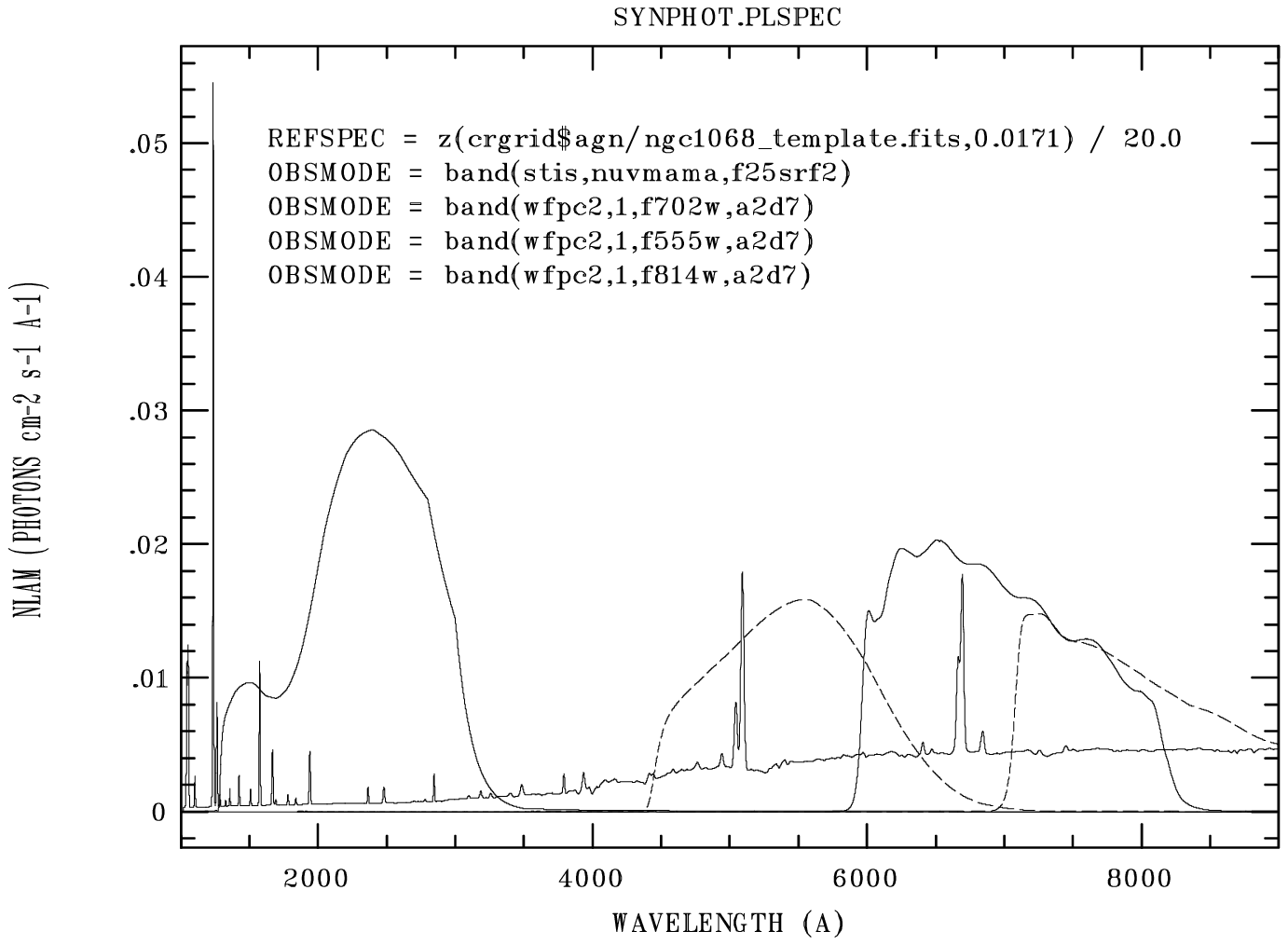,width=5in}
\psfig{figure=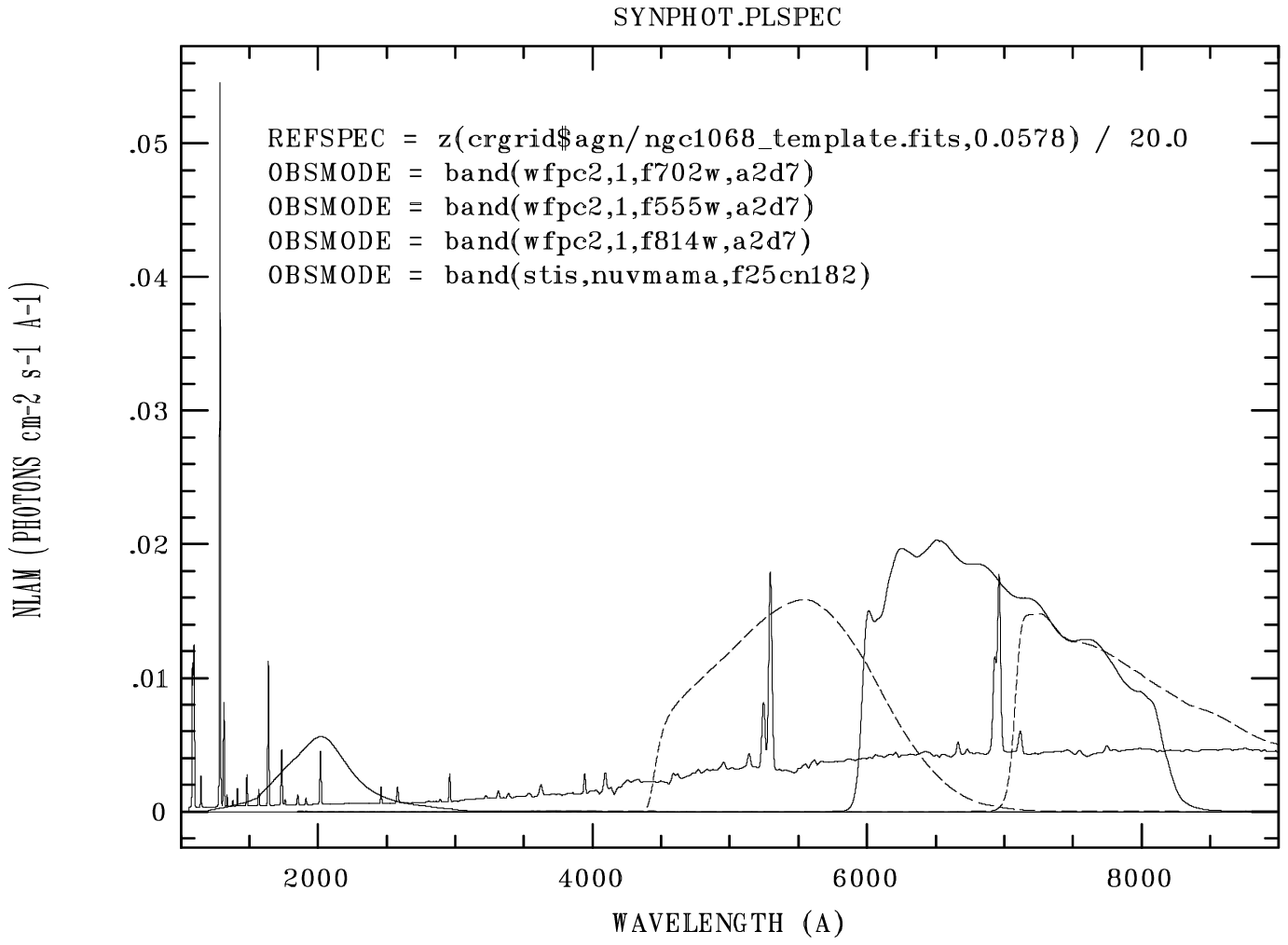,width=5in}
\figcaption[fig6a.ps,fig6b.ps]{ Band-passes of the HST STIS and WFPC2 filters. 
The STIS F25SRF2 and FCN182NM, and the WFPC2 F555W and F702W
filter bandpasses are shown. For reference these bandpasses are 
overlaid on a spectrum of NGC~1068. \label{3c449_bandpass}}

\figcaption[fig7a.ps,fig7b.ps]{3C~29: (a) STIS NUV MAMA image of 3C~29. 
The bar in the upper right corner of the image indicates the
direction of the radio jet \citep{mar99}. A scale bar in kpc at the
distance of 3C~29 is included in the lower right corner.
(b) WFPC2 image of 3C~29.  \label{3c29}}

\figcaption[fig8a.ps,fig8b.ps]{3C~35: Same as for Figure~\ref{3c29} but 
for 3C~35 \label{3c35}}

\figcaption[fig9a.ps,fig9b.ps]{3C~40: Same as for Figure~\ref{3c29} 
but for 3C~40 \label{3c40}}

\figcaption[fig10a.ps,fig10b.ps]{3C~66B: Same as for Figure~\ref{3c29} 
but for 3C~66B \label{3c66b}}

\figcaption[fig11a.ps,fig11b.ps]{3C~192: Same as for Figure~\ref{3c29}
 but for 3C~192 \label{3c192}}

\figcaption[fig12a.ps,fig12b.ps]{3C~198: Same as for Figure~\ref{3c29} 
but for 3C~198 \label{3c198}}

\figcaption[fig13a.ps,fig13b.ps]{3C~227: Same as for Figure~\ref{3c29} 
but for 3C~227 \label{3c227}}

\figcaption[fig14a.ps,fig14b.ps]{3C~231: Same as for Figure~\ref{3c29} but for 3C~231. Note that 3C~231 is not a radio galaxy so no radio axis direction
is indicated. \label{3c231}}

rightmost WF chip. \label{3c231_big}}

\figcaption[fig16a.ps,fig16b.ps]{3C~236: Same as for Figure~\ref{3c29} but 
for 3C~236 \label{3c236}}

\figcaption[fig17a.ps,fig17b.ps]{3C~270: Same as for Figure~\ref{3c29}
 but for 3C~270 \label{3c270}}

\figcaption[fig18a.ps,fig18b.ps]{3C~285: Same as for Figure~\ref{3c29} 
but for 3C~285 \label{3c285}}

\figcaption[fig19a.ps,fig19b.ps]{3C~293: Same as for Figure~\ref{3c29} 
but for 3C~293 \label{3c293}}

\figcaption[fig20a.ps,fig20b.ps]{3C~296: Same as for Figure~\ref{3c29} 
but for 3C~296 \label{3c296}}

\figcaption[fig21a.ps,fig21b.ps]{3C~305: Same as for Figure~\ref{3c29} 
but for 3C~305 \label{3c305}}

\figcaption[fig22a.ps,fig22b.ps]{3C~310: Same as for Figure~\ref{3c29} 
but for 3C~310 \label{3c310}}

\figcaption[fig23a.ps,fig23b.ps]{3C~317: Same as for Figure~\ref{3c29} 
but for 3C~317 \label{3c317}}

\figcaption[fig24a.ps,fig24b.ps]{3C~321: Same as for Figure~\ref{3c29} 
but for 3C~321 \label{3c321}}

\figcaption[fig25a.ps,fig25b.ps]{3C~326: Same as for Figure~\ref{3c29} 
but for 3C~326 \label{3c326}}

\figcaption[fig26a.ps,fig26b.ps]{3C~338: Same as for Figure~\ref{3c29} 
but for 3C~338 \label{3c338}}

\figcaption[fig27a.ps,fig27b.ps]{3C~353: Same as for Figure~\ref{3c29} 
but for 3C~353 \label{3c353}}

\figcaption[fig28a.ps,fig28b.ps]{3C~382: Same as for Figure~\ref{3c29} 
but for 3C~382 \label{3c382}}

\figcaption[fig29a.ps,fig29b.ps]{3C~388: Same as for Figure~\ref{3c29} 
but for 3C~388 \label{3c388}}

\figcaption[fig30a.ps,fig30b.ps]{3C~390.3: Same as for Figure~\ref{3c29}
 but for 3C~390.3 \label{3c390}}

\figcaption[fig31a.ps,fig31b.ps]{3C~405: Same as for Figure~\ref{3c29} 
but for 3C~405 \label{3c405}}

\figcaption[fig32a.ps,fig32b.ps]{3C~449: Same as for Figure~\ref{3c29} 
but for 3C~449 \label{3c449}}

\figcaption[fig33a.ps,fig33b.ps]{3C~452: Same as for Figure~\ref{3c29} 
but for 3C~452 \label{3c452}}

\figcaption[fig34a.ps,fig34b.ps]{3C~465: Same as for Figure~\ref{3c29}
but for 3C~465 \label{3c465}}

\clearpage

\figcaption[fig35a_1.ps,fig35a_2.ps,fig35a_3.ps,fig35a_4.ps,
            fig35a_5.ps,fig35b_1.ps,fig35b_2.ps,fig35b_3.ps,fig35b_4.ps,fig35b_5.ps,fig35b_6.ps,fig35b_7.ps,
            fig35c_1.ps,fig35c_2.ps,fig35c_3.ps,fig35c_4.ps,
            fig35d_1.ps,fig35d_2.ps,fig35d_3.ps,fig35d_4.ps,
            fig35e_1.ps,fig35e_2.ps,fig35e_3.ps
            fig35f_1.ps,fig35f_2.ps,fig35f_3.ps]{ 3C Radio Galaxies 
Categorized by UV Morphology and Scale. This figure presents all 
of the UV images of 3C Radio galaxies described in this paper. The 
images have been grouped into six categories based on their UV morphology 
and scale.  Each category is displayed along a single row in this figure, 
and the physical box size of the image is constant are displayed 
here  \label{sum_fig}}

\centerline{\hbox{
\psfig{figure=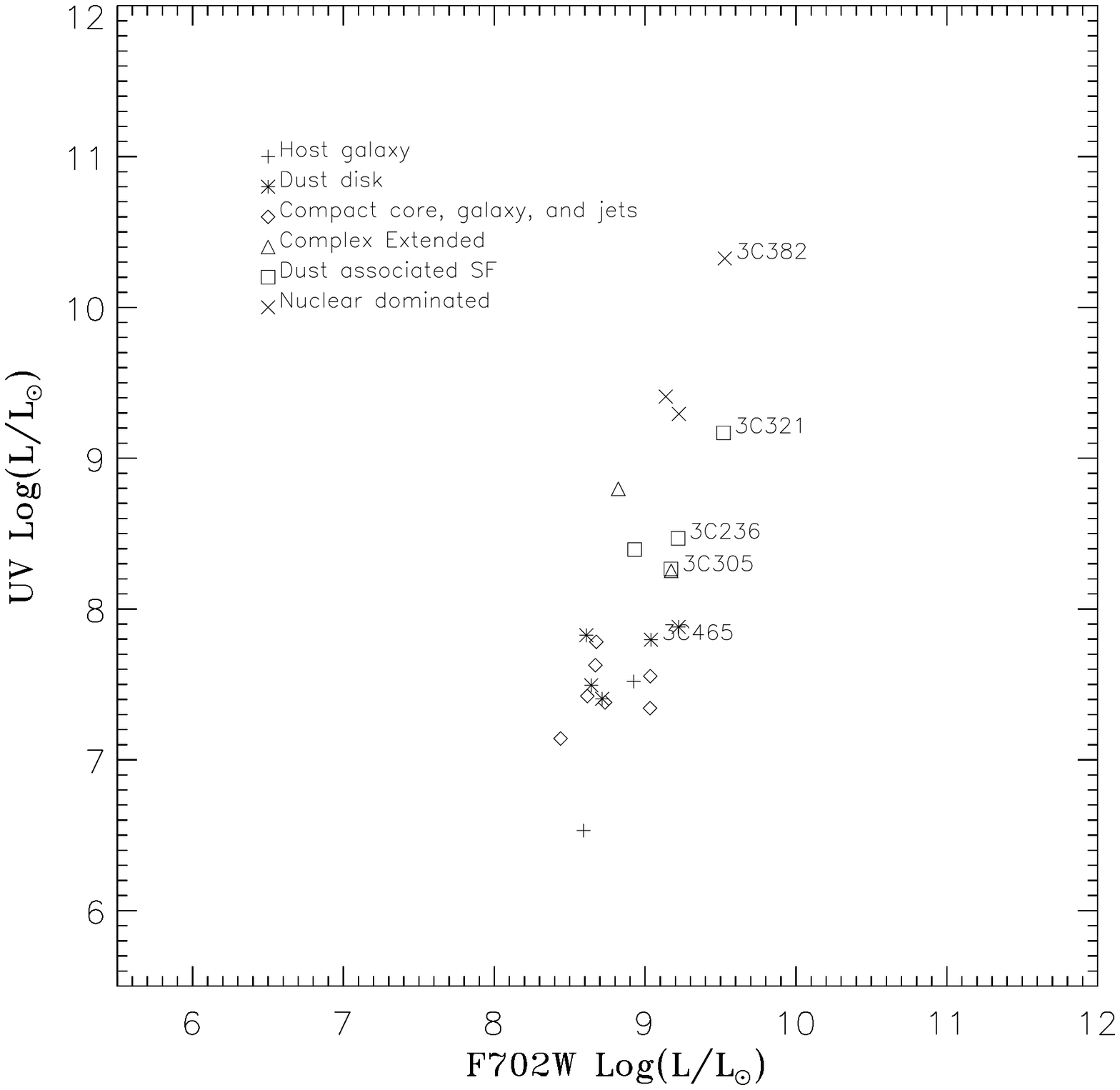,width=3.5in}
\psfig{figure=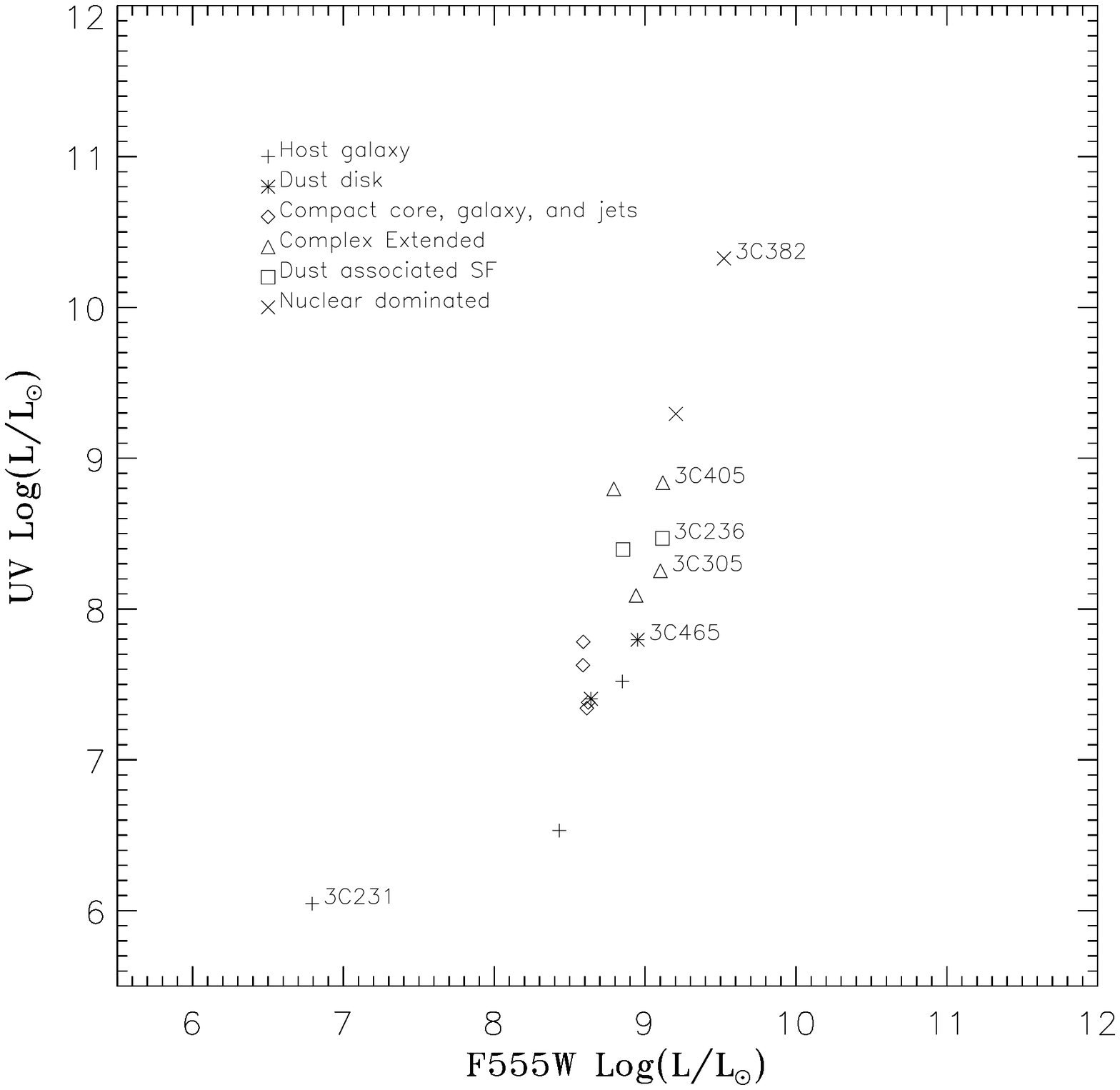,width=3.5in}}}
\figcaption[fig36a.ps,fig36b.ps]{ (a) UV luminosity versus Optical 
F702W luminosity.
(b) UV luminosity versus Optical F555W luminosity.
The plotted point styles indicate the morphological 
classifications described in section~\ref{morph_summary} \label{m5}}

\centerline{\hbox{
\psfig{figure=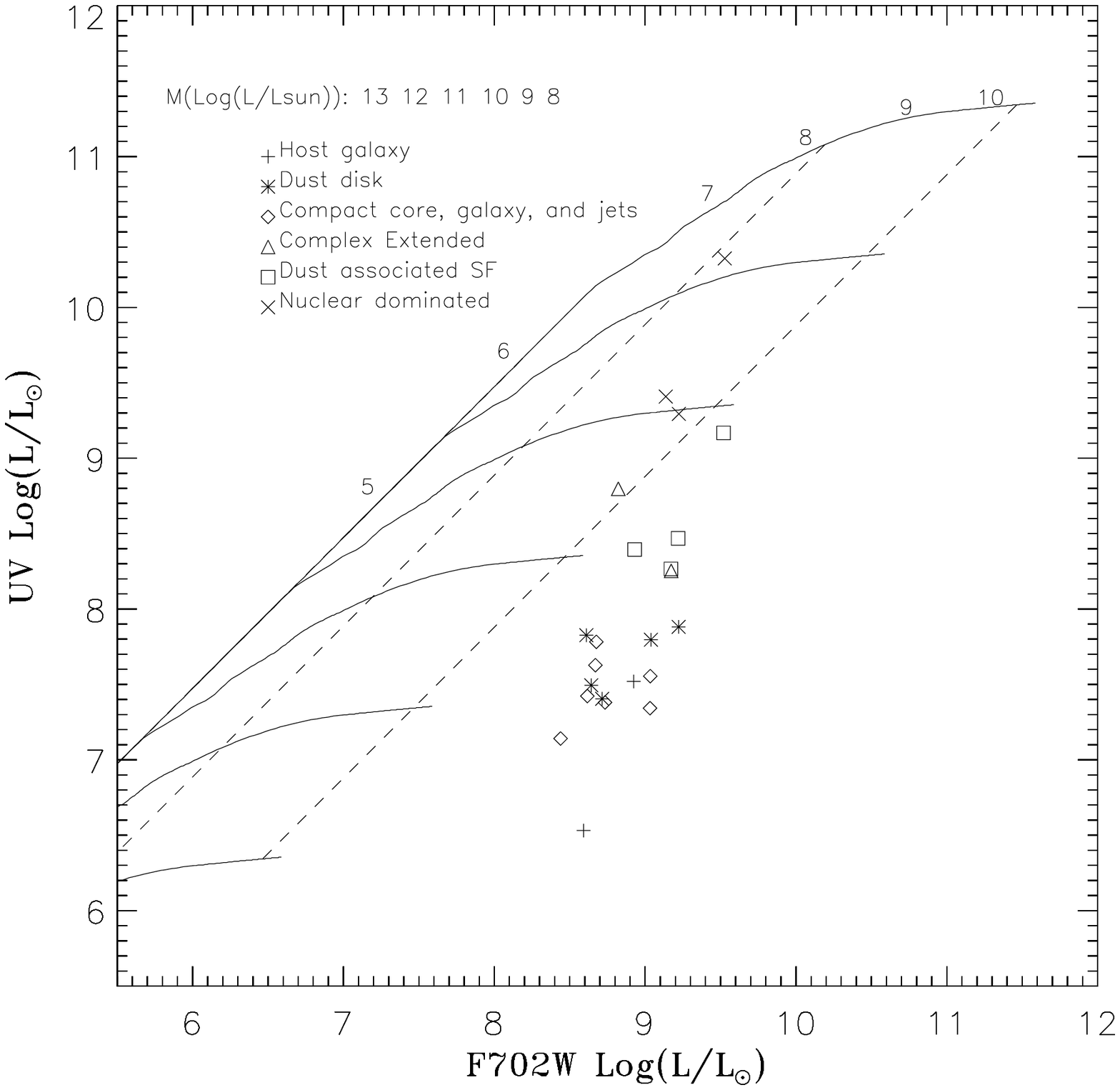,width=3.5in}
\psfig{figure=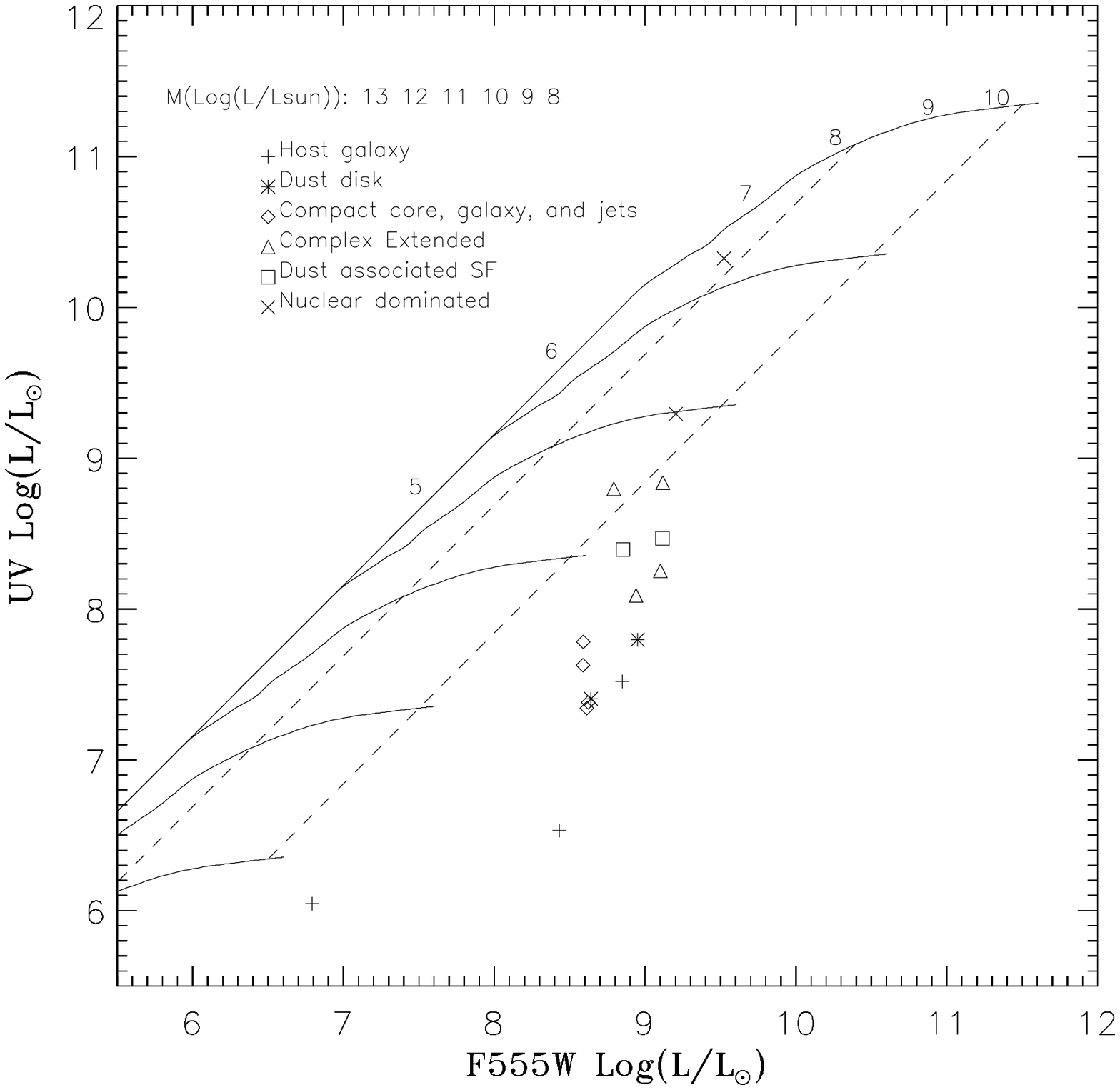,width=3.5in}}}
\figcaption[fig37a.ps,fig37b.ps]{ Constant star formation models. \label{m1}}

\centerline{\hbox{
\psfig{figure=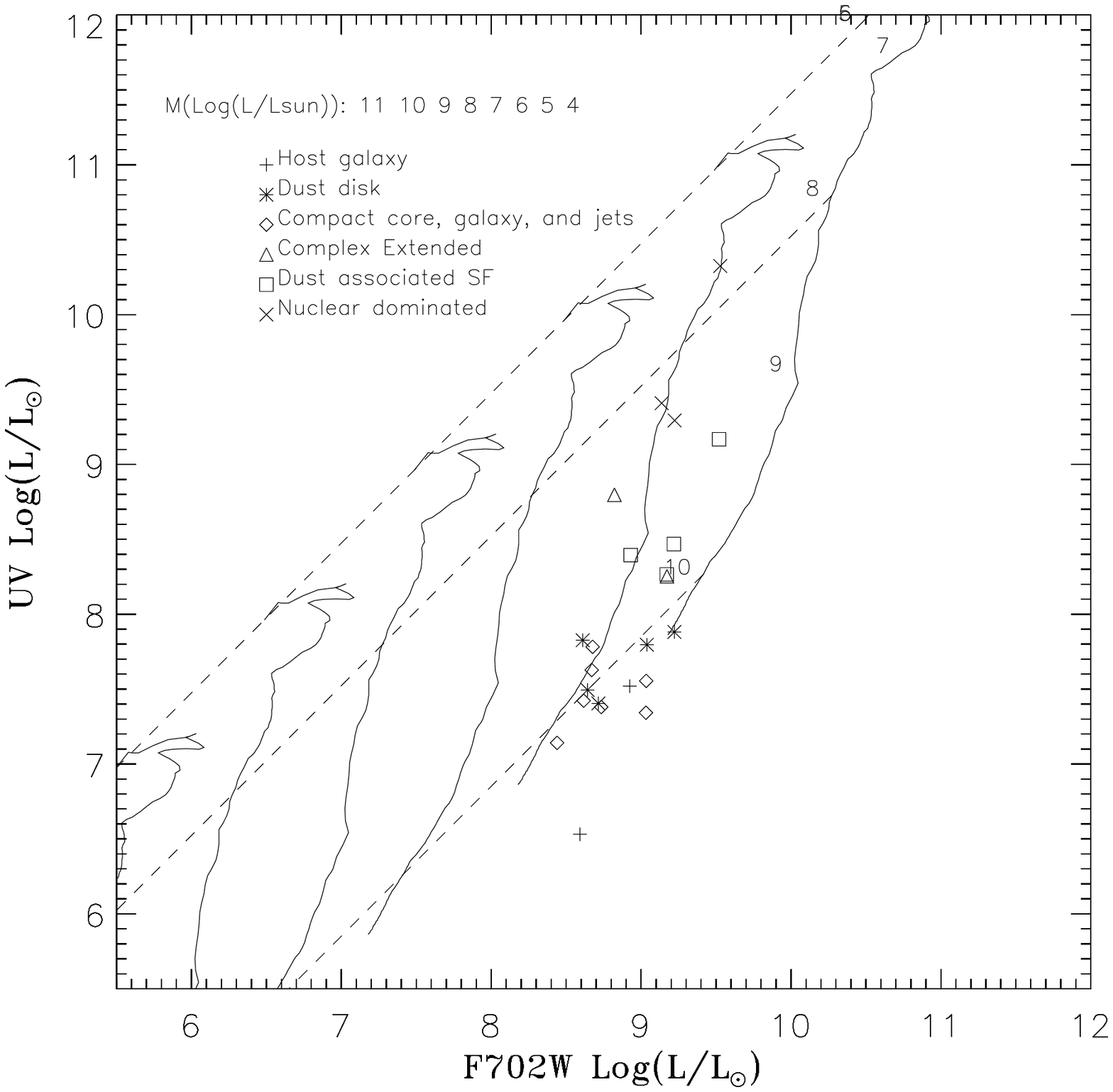,width=3.5in}
\psfig{figure=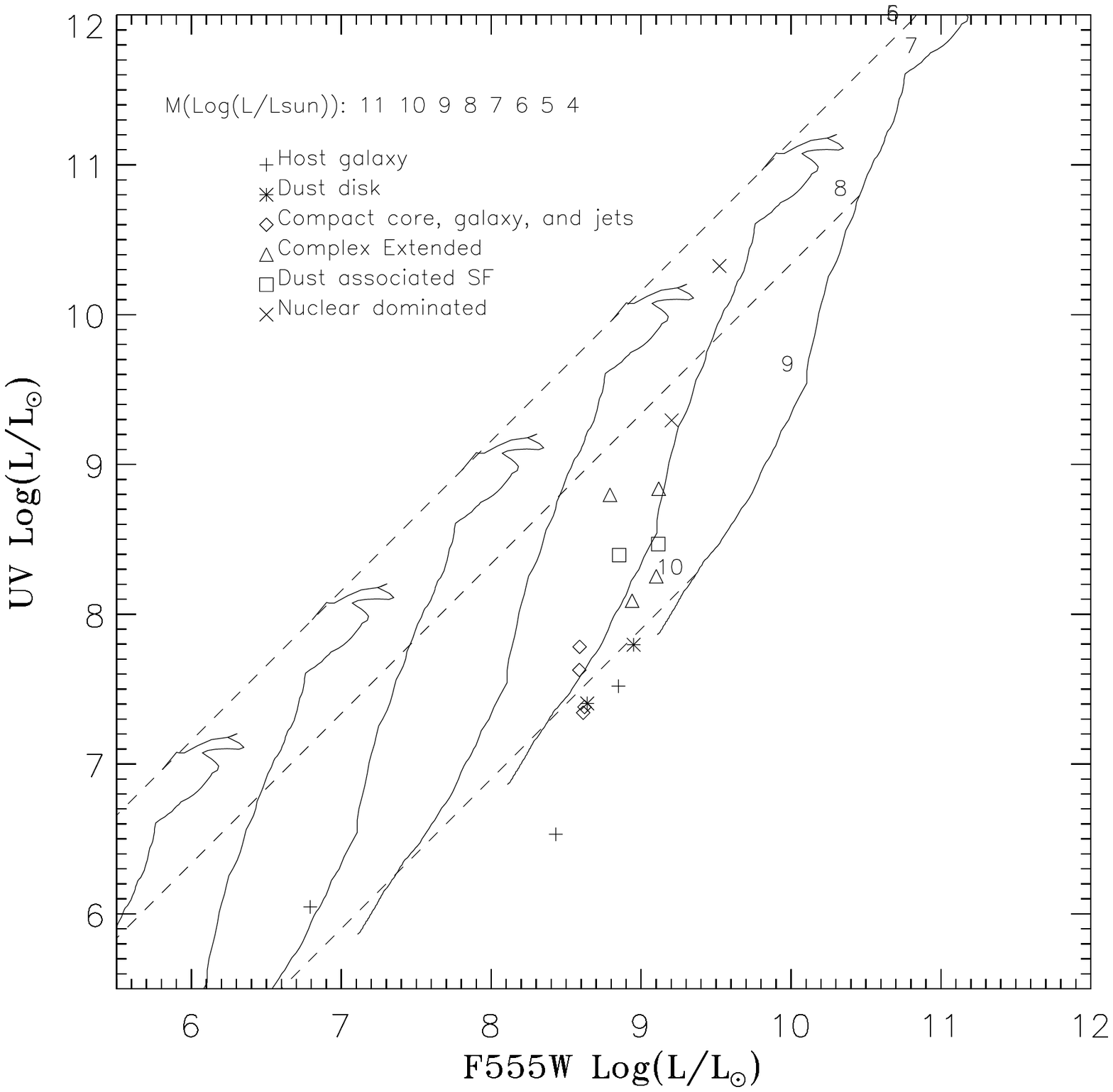,width=3.5in}}}
\figcaption[fig38a.ps,fig38b.ps]{ Single burst star formation models. 
\label{m2}}

\centerline{\hbox{
\psfig{figure=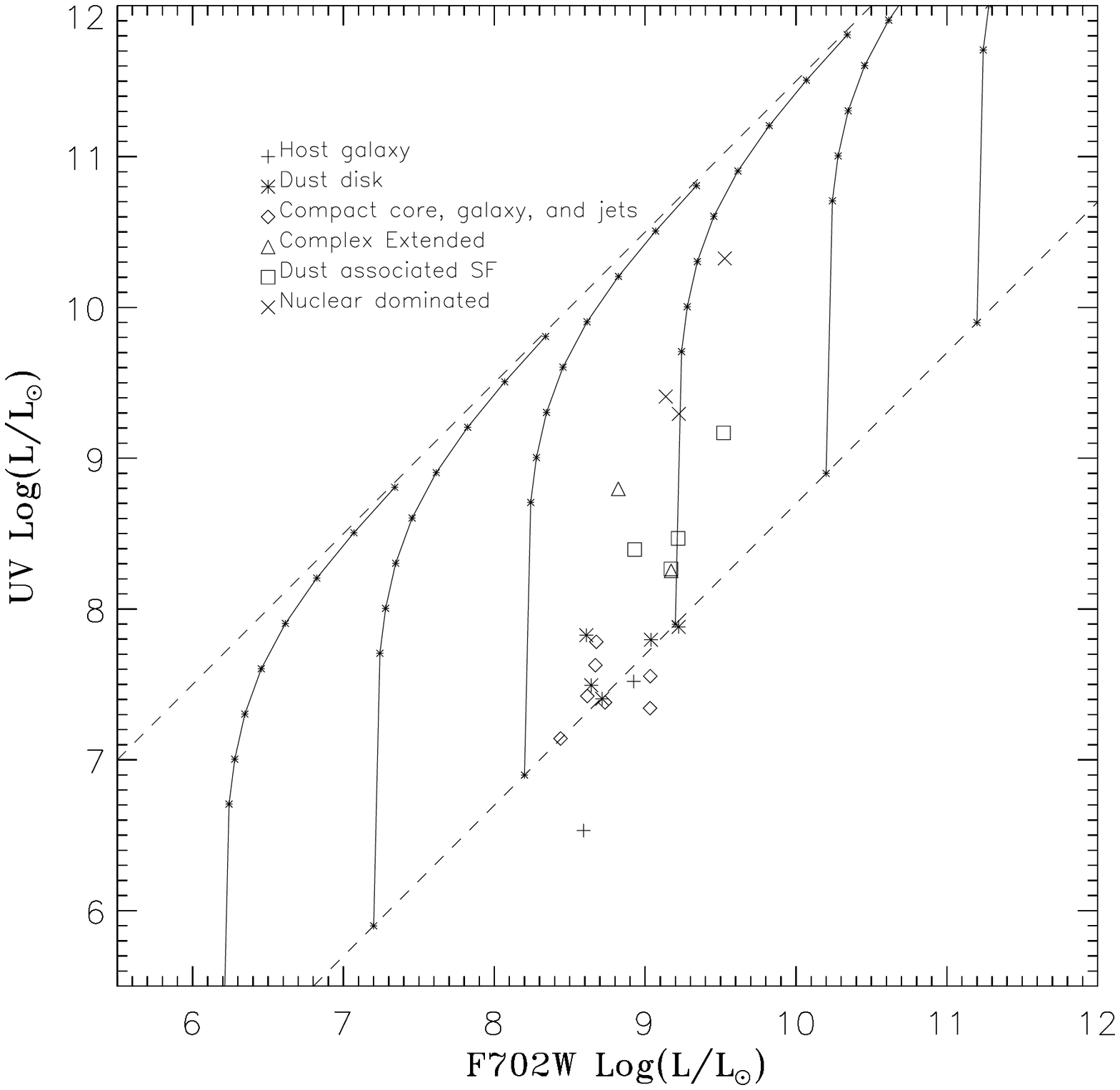,width=3.5in}
\psfig{figure=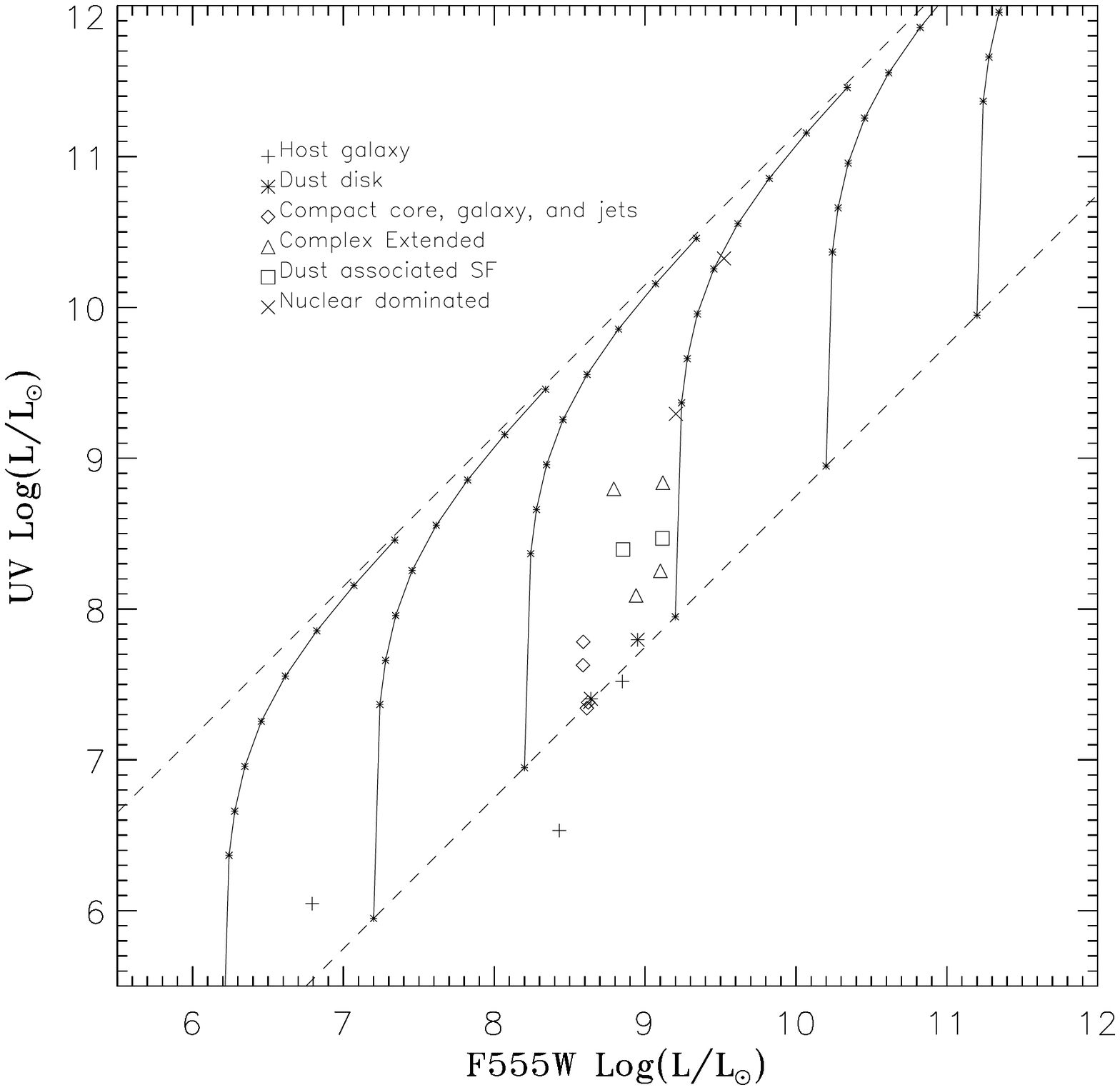,width=3.5in}}}
\figcaption[fig39a.ps,fig39b.ps]{ Young and Old starburst mixing models. 
 \label{m25}}

\centerline{\hbox{
\psfig{figure=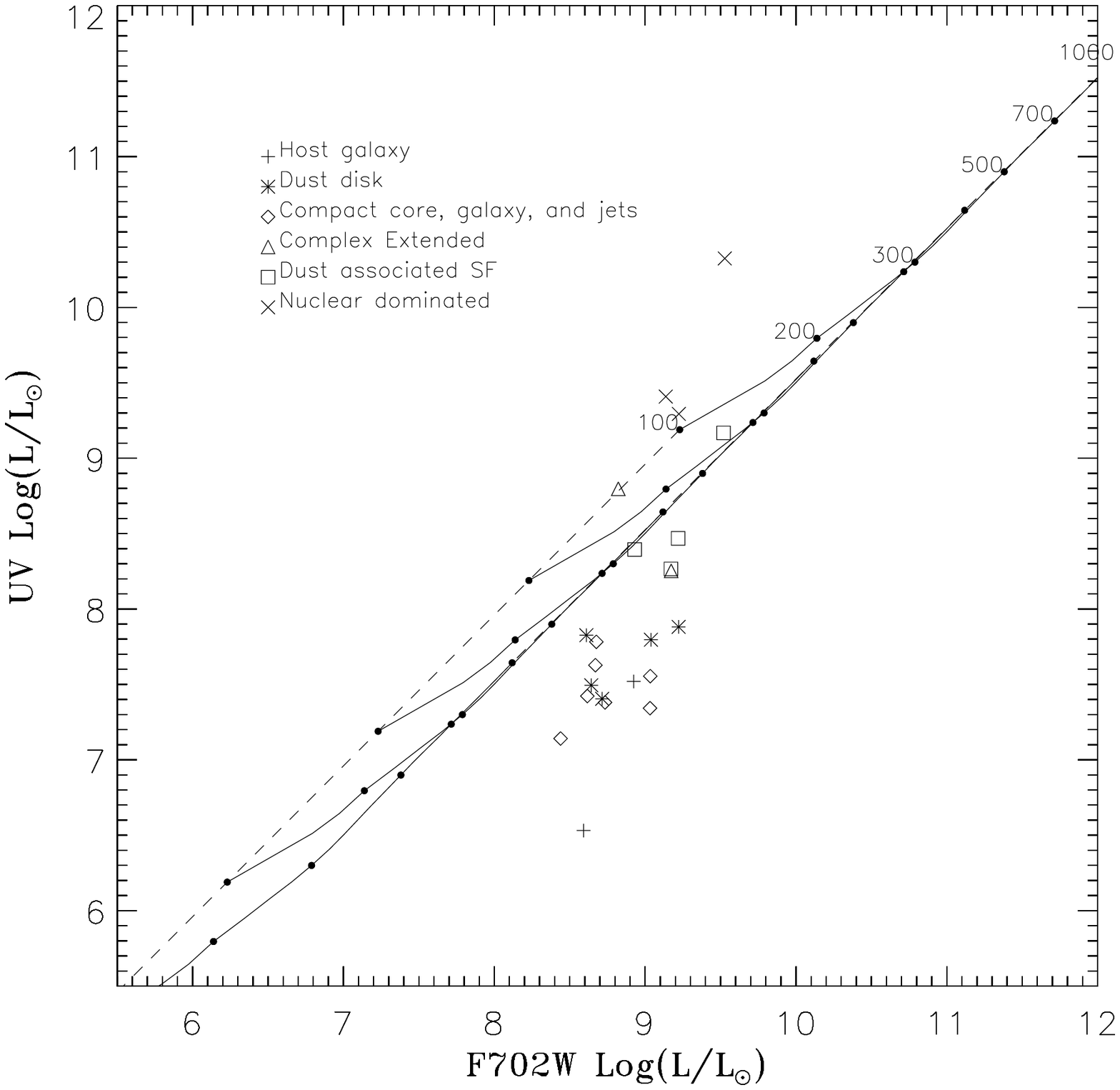,width=3.5in}
\psfig{figure=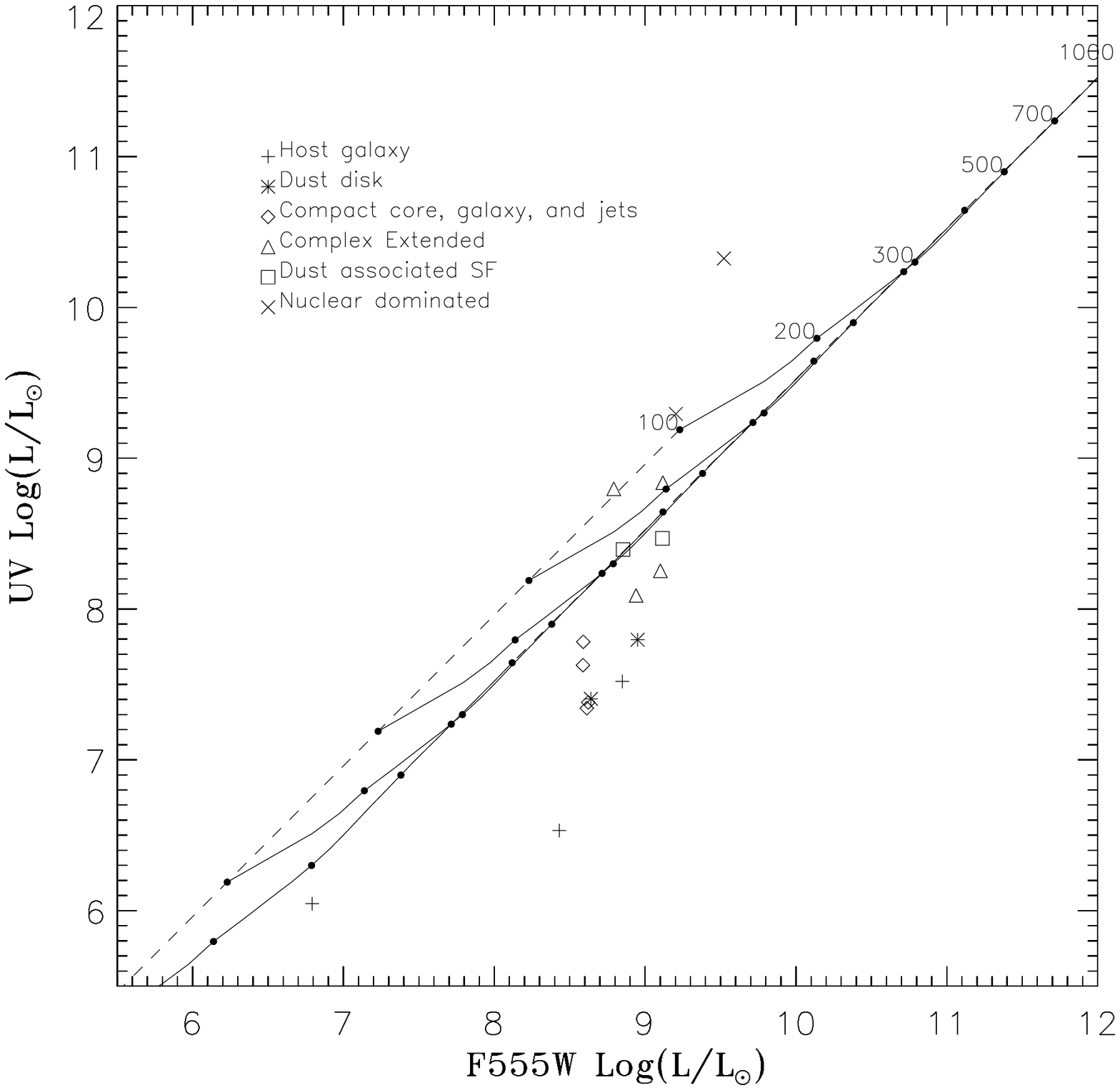,width=3.5in}}}
\figcaption[fig40a.ps,fig40b.ps]{ Shock ionization models.\label{m3}}

\centerline{\hbox{
\psfig{figure=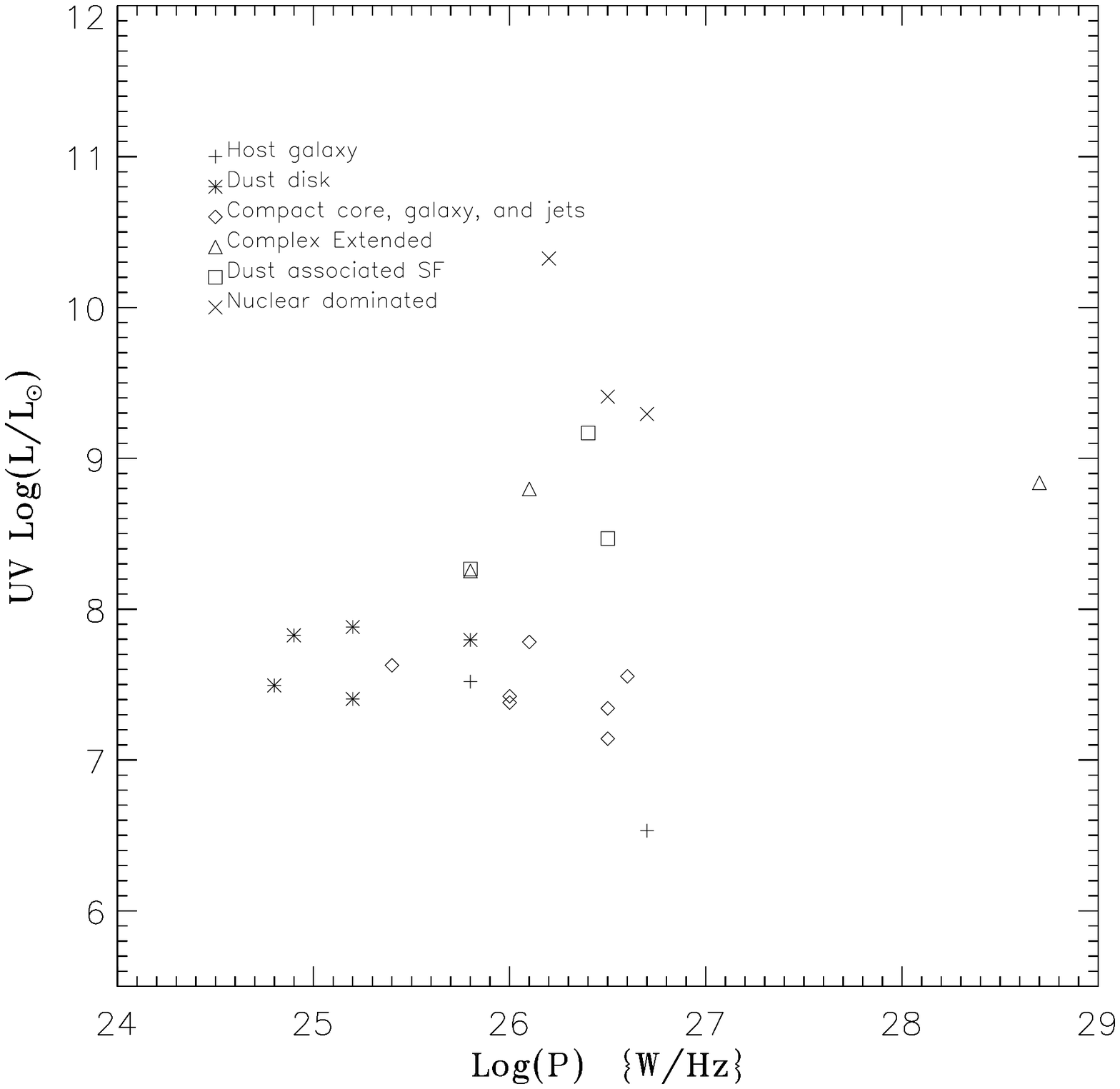,width=3.5in}
\psfig{figure=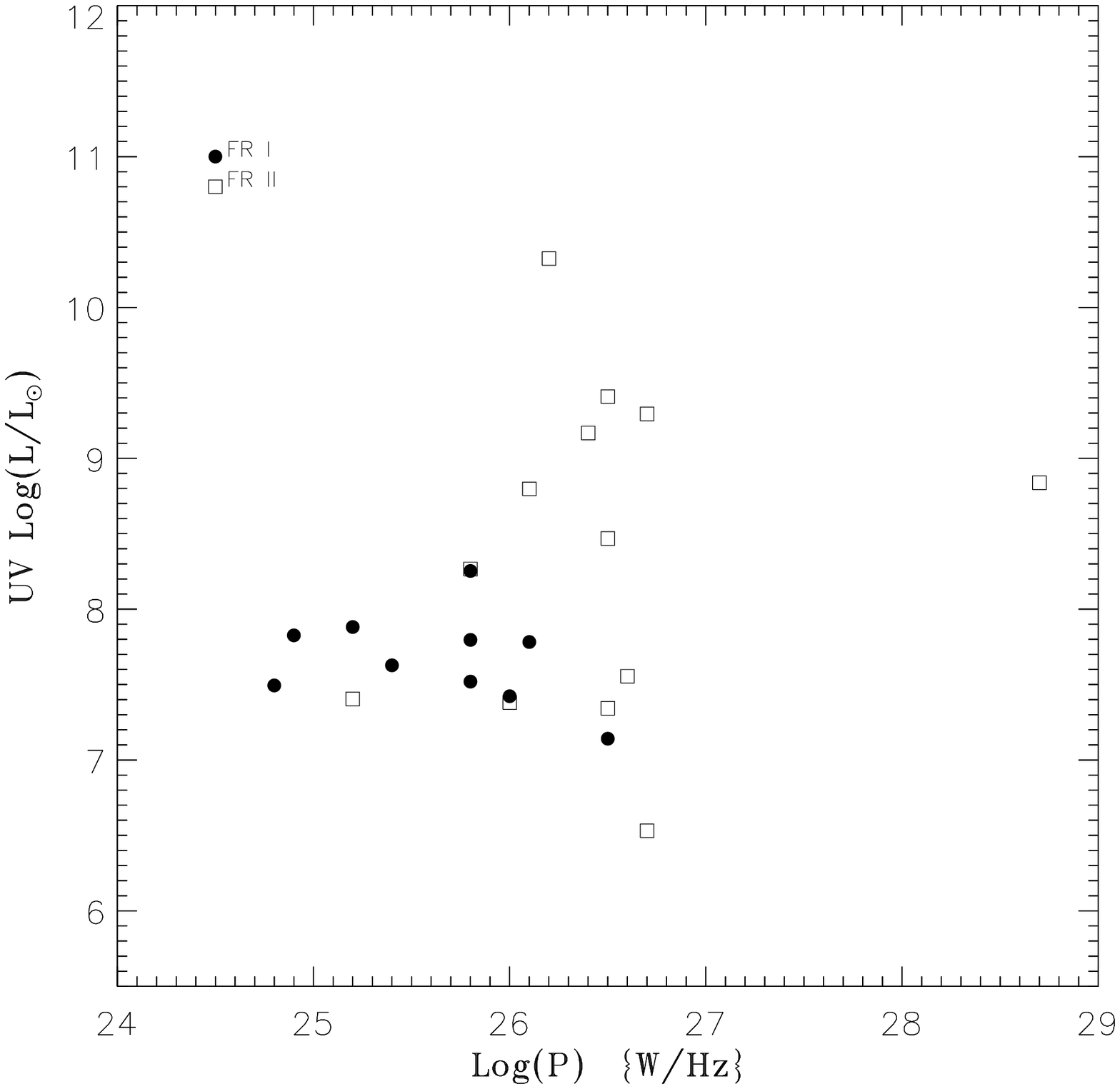,width=3.5in}}}
\figcaption[fig41a.ps,fig41b.ps]{ UV luminosity versus total radio power. 
\label{m12}}

\clearpage

\begin{deluxetable}{lcccccc}
\tablecaption{Observation Log \label{observations_table}}
\tablewidth{0pt}
\tabletypesize{\scriptsize}
\tablehead{
\colhead{} & \multicolumn{3}{c}{STIS NUV MAMA Observations} & \multicolumn{3}{c}{WFPC2 Data} \\
\colhead{3C    }       &
\colhead{STIS Observation}    &
\colhead{SRF2}         &
\colhead{F25CN182}     &
\colhead{F702W}        &
\colhead{F555W}        &
\colhead{References}  \\
\colhead{      }       &
\colhead{ Date }       &
\colhead{exposure (s)} &
\colhead{ (s)}         &
\colhead{ (s)}         &
\colhead{ (s)}         &
\colhead{      }       }
\startdata
29      & 2000 Jun 08  & 1440         &  \nodata        & 280      & 600      &   1,2     \\ 
35      & 2000 Feb 23  & 1440         &  \nodata        & 280      & 600      &   1,2     \\     
40      & 2000 Jun 03  & 1440         &  \nodata        & 560      & 600      &   1,3     \\     
66B     & 1999 Jul 13  & 1440         &  \nodata        & 280      & 460      &   1,2     \\     
192     & 2000 Mar 23  & 1440         &  \nodata        & \nodata  & 600      &   3       \\     
198     & 2000 Apr 23  & 1440         &  \nodata        & 280      & 600      &   1,2     \\     
227     & 2000 Jan 25  & 1440         &  \nodata        & 560      & 600      &   1,2     \\     
231     & 1999 Nov 10  & 1440         &  \nodata        & \nodata  & 3100     &   4       \\ 
236     & 1999 Oct 05  & 1440         &  \nodata        & 560      & 600      &   1,3     \\         
270     & 2000 Mar 05  & 1440         &  \nodata        & 280      & \nodata  &   1       \\ 
285     & 2000 Apr 16  & 1440         &  \nodata        &1200      & 600      &   1,2     \\        
293     & 2000 Jun 14  & 1440         &  \nodata        & 280      & \nodata  &   1       \\ 
296     & 2000 Apr 15  & 1440         &  \nodata        & 280      & \nodata  &   1       \\ 
305     & 2000 Apr 27  & 1440         &  \nodata        & 560      & 600      &   1,2     \\        
310     & 2000 Jun 10  & 1440         &  \nodata        & 280      & \nodata  &   1       \\ 
317     & 1999 Jul 26  & 1440         &  \nodata        & 280      & 6200     &   1,5     \\        
321     & 2000 Jun 05  & 1440         &  \nodata        & 280      & \nodata  &   1       \\ 
326     & 2000 Mar 12  & 1440         &  \nodata        & 280      & 600      &   1,2     \\        
338     & 2000 Jun 04  & 1440         &  \nodata        & 280      & \nodata  &   1       \\ 
353     & 2000 Jun 22  & 1440         &  \nodata        & 280      & 600      &   1,2     \\        
382     & 2000 Feb 23  & \nodata      &  1440           & 280      & 600      &   1,2     \\        
388     & 2000 Jun 02  & 1440         & \nodata         & 280      & \nodata  &   1       \\ 
390.3   & 1999 Aug 10  & \nodata      &  1440           & 280      & \nodata  &   1       \\ 
405     & 2000 Jun 25  & 2160         & \nodata         &          & 2700     &   6       \\      
449     & 2000 Apr 16  & 1440         & \nodata         & 560      & \nodata  &   1       \\ 
452     & 2000 Jan 30  & \nodata      &  1440           & 280      & \nodata  &   1       \\ 
465     & 2000 Jan 30  & 1440         & \nodata         & 280      & 600      &   1,2     \\ 
\enddata
\tablecomments{ References: HST program identification and principal investigator 
for F702W and F555W data: 1-- GO~5476 Sparks, 2-- GO~6967 Sparks, 3-- GO~6348 Sparks,
4-- GO~7446 O'Connell, 5-- GO~6810 Geisler, 6-- GO~5368 Jackson }
\end{deluxetable}

\begin{deluxetable}{lcccccc}
\tablecaption{Radio Properties of the STIS snapshot sample. \label{radio_prop_table} }
\tablewidth{0pt}
\tabletypesize{\scriptsize}
\tablehead{   
\colhead{3C }            &
\colhead{z }             &
\colhead{$S_{178}$}      &
\colhead{log~$P_{178}$ } &
\colhead{$\alpha$}       &
\colhead{FR}             &
\colhead{E(B-V)} }
\startdata
29      &   0.0447   &  15.1  &  25.8 &  0.50 &  I    &     0.036     \\
35      &   0.0670   &  10.5  &  26.0 &  0.77 &  II   &     0.141      \\
40      &   0.0177   &  26.0  &  25.2 &  0.66 &  II    &     0.041     \\
66b     &   0.0215   &  24.6  &  25.4 &  0.62 &  I    &     0.080     \\
192     &   0.0598   &  21.1  &  \nodata     &  0.79 &  II   &     0.054      \\
198     &   0.0815   &  9.7   &  26.1 &  0.69 &  II   &     0.026      \\
227     &   0.0861   &  30.4  &  26.7 &  0.67 &  II   &     0.026      \\
231     &   0.000677 &  \nodata      &  \nodata     &  \nodata     & \nodata      &     0.159  \\
236     &   0.0989   &  14.4  &  26.5 &  0.51 &  II   &     0.011      \\
270     &   0.0073   &  51.8  &  24.8 &  0.51 &  I    &     0.018     \\
285     &   0.0794   &  11.3  & \nodata  &  0.95 &  II   &     0.017      \\
293     &   0.0452   &  12.7  &  25.8 &  0.45 &  II   &     0.017      \\
296     &   0.0237   &  13.0  &  25.2 &  0.67 &  I    &     0.025     \\
305     &   0.041439 &  15.7  &  25.8 &  0.85 &  I    &     0.029      \\
310     &   0.0540   &  55.1  &  26.5 &  0.92 &  I   &     0.042      \\
317     &   0.0350   &  49.0  &  26.1 &  1.02 &  I    &     0.037     \\
321     &   0.0960   &  13.5  &  26.4 &  0.60 &  II   &     0.044      \\
326     &   0.0885   &  20.4  &  26.5 &  0.88 &  II   &     0.053      \\
338     &   0.0298   &  46.9  &  26.0 &  1.19 &  I    &     0.012     \\
353     &   0.0304   & 236.0  &  26.7 &  0.71 &  II   &     0.439      \\
382     &   0.0578   &  19.9  &  26.2 &  0.59 &  II   &     0.070      \\
388     &   0.0908   &  24.6  &  26.6 &  0.70 &  II   &     0.080      \\
390.3   &   0.0561   &  47.5  &  26.5 &  0.75 &  II   &     0.071      \\
405     &   0.056075 &  8700  &  28.7 &  0.74 &  II   &     0.381      \\
449     &   0.0171   &  11.5  &  24.9 &  0.58 &  I    &     0.167     \\
452     &   0.0811   &  54.4  &  26.9 &  0.78 &  II   &     0.137      \\
465     &   0.0293   &  37.8  &  25.8 &  0.75 &  I    &     0.069     \\
\enddata
\end{deluxetable}

\begin{deluxetable}{lccc}
\tablecaption{ F$_{\lambda}$ Measurements. \label{tfb2_flambda_table}}
\tablewidth{0pt}
\tabletypesize{\scriptsize}
\tablehead{   
\colhead{3C }         &
\colhead{UV }         &
\colhead{F702W }      &
\colhead{F555W }      \\
\colhead{ }       &
\colhead{(erg~s$^{-1}$~cm$^{-2}$~\AA$^{-1}$) } &
\colhead{(erg~s$^{-1}$~cm$^{-2}$~\AA$^{-1}$) } &
\colhead{(erg~s$^{-1}$~cm$^{-2}$~\AA$^{-1}$) }} 
\startdata
29      &  5.272$\times10^{17}$ $\pm$ 1.1$\times10^{18}$  &  1.333$\times10^{15}$ $\pm$ 3.5$\times10^{18}$  &  1.223$\times10^{15}$ $\pm$ 3.1$\times10^{18}$  \\
35      &  8.123$\times10^{18}$ $\pm$ 3.2$\times10^{19}$  &  3.042$\times10^{16}$ $\pm$ 1.5$\times10^{18}$  &  2.376$\times10^{16}$ $\pm$ 1.2$\times10^{18}$  \\
40      &  2.489$\times10^{16}$ $\pm$ 1.5$\times10^{18}$  &  5.184$\times10^{15}$ $\pm$ 4.7$\times10^{18}$  &  4.762$\times10^{15}$ $\pm$ 6.1$\times10^{18}$  \\
66b     &  2.138$\times10^{16}$ $\pm$ 1.2$\times10^{18}$  &  2.915$\times10^{15}$ $\pm$ 5.2$\times10^{18}$  &  2.550$\times10^{15}$ $\pm$ 5.2$\times10^{18}$  \\
192     &  9.636$\times10^{17}$ $\pm$ 9.1$\times10^{19}$  &        \nodata                                  &  7.995$\times10^{16}$ $\pm$ 2.4$\times10^{18}$  \\
198     &  3.228$\times10^{16}$ $\pm$ 1.4$\times10^{18}$  &  3.235$\times10^{16}$ $\pm$ 1.7$\times10^{18}$  &  3.329$\times10^{16}$ $\pm$ 1.6$\times10^{18}$  \\
227     &  9.069$\times10^{16}$ $\pm$ 2.0$\times10^{18}$  &  7.311$\times10^{16}$ $\pm$ 1.7$\times10^{18}$  &  7.697$\times10^{16}$ $\pm$ 2.4$\times10^{18}$  \\
231     &  3.250$\times10^{15}$ $\pm$ 5.4$\times10^{18}$  &        \nodata                                  &  3.266$\times10^{14}$ $\pm$ 8.1$\times10^{18}$  \\
236     &  5.667$\times10^{17}$ $\pm$ 8.1$\times10^{19}$  &  4.556$\times10^{16}$ $\pm$ 1.4$\times10^{18}$  &  3.705$\times10^{16}$ $\pm$ 1.7$\times10^{18}$  \\
270     &  2.120$\times10^{15}$ $\pm$ 3.4$\times10^{18}$  &  2.718$\times10^{14}$ $\pm$ 1.7$\times10^{17}$  &  \nodata                                        \\
285     &  1.435$\times10^{16}$ $\pm$ 9.6$\times10^{19}$  &  4.466$\times10^{16}$ $\pm$ 9.7$\times10^{19}$  &  4.144$\times10^{16}$ $\pm$ 1.9$\times10^{18}$  \\
293     &  3.291$\times10^{16}$ $\pm$ 2.2$\times10^{18}$  &  2.397$\times10^{15}$ $\pm$ 5.1$\times10^{18}$  &  \nodata                                        \\
296     &  4.668$\times10^{16}$ $\pm$ 1.6$\times10^{18}$  &  9.637$\times10^{15}$ $\pm$ 8.2$\times10^{18}$  &  \nodata                                        \\
305     &  3.490$\times10^{16}$ $\pm$ 1.4$\times10^{18}$  &  2.787$\times10^{15}$ $\pm$ 3.4$\times10^{18}$  &  2.599$\times10^{15}$ $\pm$ 4.5$\times10^{18}$  \\
310     &  1.448$\times10^{17}$ $\pm$ 3.2$\times10^{19}$  &  2.949$\times10^{16}$ $\pm$ 1.5$\times10^{18}$  &  \nodata                                        \\
317     &  1.563$\times10^{16}$ $\pm$ 1.2$\times10^{18}$  &  1.225$\times10^{15}$ $\pm$ 3.9$\times10^{18}$  &  1.097$\times10^{15}$ $\pm$ 1.1$\times10^{18}$  \\
321     &  4.798$\times10^{16}$ $\pm$ 1.6$\times10^{18}$  &  1.120$\times10^{15}$ $\pm$ 3.1$\times10^{18}$  &  \nodata                                        \\
326     &  7.919$\times10^{18}$ $\pm$ 3.2$\times10^{19}$  &  4.205$\times10^{16}$ $\pm$ 1.8$\times10^{18}$  &  1.727$\times10^{16}$ $\pm$ 1.0$\times10^{18}$  \\
338     &  1.127$\times10^{16}$ $\pm$ 9.6$\times10^{19}$  &  1.556$\times10^{15}$ $\pm$ 4.2$\times10^{18}$  &  \nodata                                        \\
353     &  7.511$\times10^{19}$ $\pm$ 4.3$\times10^{19}$  &  5.578$\times10^{16}$ $\pm$ 2.0$\times10^{18}$  &  3.109$\times10^{16}$ $\pm$ 1.5$\times10^{18}$  \\
382     &  2.594$\times10^{14}$ $\pm$ 3.4$\times10^{17}$  &  2.970$\times10^{15}$ $\pm$ 4.6$\times10^{18}$  &  3.127$\times10^{15}$ $\pm$ 4.4$\times10^{18}$  \\
388     &  1.013$\times10^{17}$ $\pm$ 5.2$\times10^{19}$  &  3.776$\times10^{16}$ $\pm$ 1.8$\times10^{18}$  &  \nodata                                        \\
390.3   &  3.320$\times10^{15}$ $\pm$ 1.3$\times10^{17}$  &  1.278$\times10^{15}$ $\pm$ 3.0$\times10^{18}$  &  \nodata                                        \\
405     &  6.536$\times10^{17}$ $\pm$ 6.0$\times10^{19}$  &        \nodata                                  &  5.255$\times10^{16}$ $\pm$ 8.6$\times10^{19}$  \\
449     &  2.906$\times10^{16}$ $\pm$ 1.9$\times10^{18}$  &  3.317$\times10^{15}$ $\pm$ 3.7$\times10^{18}$  &  \nodata                                        \\
465     &  1.834$\times10^{16}$ $\pm$ 1.6$\times10^{18}$  &  3.753$\times10^{15}$ $\pm$ 5.7$\times10^{18}$  &  3.264$\times10^{15}$ $\pm$ 5.0$\times10^{18}$  \\
\enddata
\end{deluxetable}

\begin{deluxetable}{lccc}
\tablecaption{Luminosity Measurements. \label{tfb2_lum_table}}
\tablewidth{0pt}
\tabletypesize{\scriptsize}
\tablehead{   
\colhead{3C }       &
\colhead{NUV }      &         
\colhead{F702W }    &
\colhead{F555W }    \\
\colhead{ }           &
\colhead{log(L/L$_{\sun}$)} &
\colhead{log(L/L$_{\sun}$)} &
\colhead{log(L/L$_{\sun}$)}}
\startdata
  29    &      7.520 $\pm$    0.12  &      8.924 $\pm$    0.04  &      8.849 $\pm$    0.05  \\
  35    &      7.380 $\pm$    0.45  &      8.734 $\pm$    0.14  &      8.623 $\pm$    0.18  \\
  40    &      7.404 $\pm$    0.13  &      8.715 $\pm$    0.04  &      8.641 $\pm$    0.05  \\
  66B   &      7.628 $\pm$    0.25  &      8.670 $\pm$    0.08  &      8.588 $\pm$    0.10  \\
 192    &      8.090 $\pm$    0.17  &    \nodata                &      8.940 $\pm$    0.07  \\
 198    &      8.797 $\pm$    0.08  &      8.822 $\pm$    0.03  &      8.792 $\pm$    0.04  \\
 227    &      9.294 $\pm$    0.08  &      9.224 $\pm$    0.03  &      9.204 $\pm$    0.03  \\
 231    &      6.046 $\pm$    0.49  &    \nodata                &      6.793 $\pm$    0.20  \\
 236    &      8.468 $\pm$    0.35  &      9.218 $\pm$    0.11  &      9.114 $\pm$    0.14  \\
 270    &      7.494 $\pm$    0.06  &      8.643 $\pm$    0.02  &    \nodata                \\
 285    &      8.395 $\pm$    0.06  &      8.931 $\pm$    0.02  &      8.853 $\pm$    0.02  \\
 293    &      8.266 $\pm$    0.06  &      9.171 $\pm$    0.02  &    \nodata                \\
 296    &      7.882 $\pm$    0.08  &      9.222 $\pm$    0.02  &    \nodata                \\
 305    &      8.253 $\pm$    0.09  &      9.172 $\pm$    0.03  &      9.101 $\pm$    0.04  \\
 310    &      7.141 $\pm$    0.14  &      8.439 $\pm$    0.04  &    \nodata                \\
 317    &      7.782 $\pm$    0.12  &      8.676 $\pm$    0.04  &      8.590 $\pm$    0.05  \\
 321    &      9.167 $\pm$    0.14  &      9.520 $\pm$    0.04  &    \nodata                \\
 326    &      7.342 $\pm$    0.18  &      9.032 $\pm$    0.05  &      8.613 $\pm$    0.07  \\
 338    &      7.423 $\pm$    0.04  &      8.617 $\pm$    0.01  &    \nodata                \\
 353    &      6.531 $\pm$    1.55  &      8.592 $\pm$    0.42  &      8.431 $\pm$    0.56  \\
 382    &     10.324 $\pm$    0.22  &      9.528 $\pm$    0.07  &      9.523 $\pm$    0.09  \\
 388    &      7.555 $\pm$    0.27  &      9.034 $\pm$    0.08  &    \nodata                \\
 390.3  &      9.409 $\pm$    0.22  &      9.137 $\pm$    0.07  &    \nodata                \\
 405    &      8.838 $\pm$    1.14  &    \nodata                &      9.117 $\pm$    0.49  \\
 449    &      7.826 $\pm$    0.51  &      8.610 $\pm$    0.16  &    \nodata                \\
 465    &      7.796 $\pm$    0.22  &      9.039 $\pm$    0.07  &      8.950 $\pm$    0.09  \\
\enddata
\end{deluxetable}

\end{document}